\documentclass[10pt]{iopart}

\usepackage{iopams}  
\usepackage[superscript,biblabel]{cite}
\usepackage[colorlinks]{hyperref}\hypersetup{
     colorlinks   = true,
     citecolor    = blue
}
\usepackage{hyperref}
\hypersetup{
    colorlinks=true,
    linkcolor=blue,
    filecolor=magenta,      
    urlcolor=cyan,
}
\usepackage[normalem]{ulem}
\usepackage[toc,page]{appendix}
\usepackage[utf8]{inputenc}
\usepackage[T1]{fontenc}
\usepackage{lmodern}
\usepackage[english]{babel}
\usepackage{color}
\usepackage{graphics}
\usepackage{epsfig}
\usepackage{float}
\usepackage{array}
\usepackage{bm}
\usepackage{footnote}
\usepackage{gensymb}
\usepackage[symbol]{footmisc}
\usepackage{array}
\usepackage{makecell}
\usepackage{tabularx}
\usepackage{pdfpages}
\expandafter\let\csname equation*\endcsname\relax
\expandafter\let\csname endequation*\endcsname\relax
\usepackage{amsmath}

\newcolumntype{C}[1]{>{\centering\arraybackslash$}p{#1}<{$}}
\usepackage{fancyhdr}
\usepackage{lastpage}

\usepackage[heightrounded]{geometry}
\usepackage{iopams} 

\begin{document}

\setlength{\headwidth}{\textwidth}
\pagestyle{fancyplain}
%\fancyhf{}
\lhead{ \fancyplain{}{Nucl. Fusion XX (2026) XXXXXX} }
\rhead{ \fancyplain{}{A. Kumar \it et  al} }
%\rfoot{Page \thepage \hspace{1pt} of \pageref{LastPage}}
\setlength{\headwidth}{\textwidth}
\addtolength{\headwidth}{\marginparsep}
\addtolength{\headwidth}{\marginparwidth}

\title[]{RF-Specific Tungsten Erosion and Global Transport in ITER under Neon Seeding\mbox{*}}

\author{
Atul Kumar$^{1\dagger}$,
Dhyanjyoti Nath$^{2}$,
Wouter Tierens$^{1}$,
Jeremy D. Lore$^{1}$,
Andrei Pshenov$^{3}$,
Tom Wauters$^{3}$,
Andrea Galvan$^{4}$,
Davide Curreli$^{4}$,
Syun'ichi Shiraiwa$^{5}$,
Nicola Bertelli$^{5}$,
Guillaume Urbanczyk$^{6}$,
Sebastijan Brezinsek$^{7}$,
Onkar Sahni$^{2}$,
Mark Shephard$^{2}$,
Walid Helou$^{3}$,
Xavier Bonnin$^{3}$,
Klaus Schmid$^{8}$,
Volodymyr Bobkov$^{8}$
}

\address{$^1$Oak Ridge National Laboratory, PO Box 2008, Mail Stop 6305, Bldg. 7601, Oak Ridge, TN 37831, USA}

\address{$^2$Rensselaer Polytechnic Institute, Troy, NY 12180, USA}

\address{$^3$ITER Organization, Route de Vinon-sur-Verdon, CS 90 046, 13067 Saint-Paul-lez-Durance Cedex, France}

\address{$^4$University of Illinois Urbana-Champaign, Urbana, IL 61801, USA}
\address{$^5$Princeton Plasma Physics Laboratory, Princeton, NJ 08536, USA }

\address{$^6$Institut Jean Lamour, UMR 7198 CNRS--Université de Lorraine, 2 Allée André Guinier, F-54011 Nancy, France}

\address{$^7$Institut für Plasmaphysik, Forschungszentrum Jülich GmbH, 52425 Jülich, Germany}
\address{$^8$Max-Planck-Institut für Plasmaphysik, Boltzmannstr.~2, 85748 Garching, Germany}

% \address{$^4$University of Wisconsin, Madison, WI 53706, USA}
% \address{$^5$University of Illinois at Urbana Champaign, Urbana, IL 61801, USA }

% \address{$^a$see http://west.cea.fr/WESTteam }

% \address{$^1$Oak Ridge National Laboratory, 1 Bethel Valley Road, Oak Ridge, TN-37831, USA}
% \address{$^2$Rensselaer Polytechnic Institute, Troy, NY,  USA}
% \address{$^3$General Atomics, San Diego,  CA, USA }
% \address{$^4$University of Wisconsin, Madison, WI 53706, USA}
% \address{$^5$University of Illinois at Urbana Champaign, Urbana, IL 61801, USA }
% \address{$^6$Princeton Plasma Physics Laboratory, Princeton, NJ 08536, USA }
% \address{$^a$see http://west.cea.fr/WESTteam }

\ead{$^\dagger$kumara@ornl.gov}

% \vspace{10pt}
% \begin{indented}
%  \item[]{March 2021}
% \end{indented}

\begin{abstract}
Ion cyclotron radio-frequency heating (ICRH) is a key
auxiliary heating system in ITER, but high-power RF
operation can enhance plasma--material interactions
through rectified RF sheath potentials on antenna
structures and nearby plasma-facing components. We
present the first predictive application of the STRIPE
(\textit{Simulated Transport of RF Impurity Production
and Emission}) framework to investigate RF
sheath-driven tungsten (W) erosion and global impurity
transport from the ITER ICRH antenna under
ITER-relevant Ne-seeded conditions. The framework
couples wide-grid \textit{SOLPS-ITER} plasma
backgrounds, full-wave RF sheath calculations,
geometry-specific ion energy--angle distributions
(IEADs), sputtering physics, and three-dimensional
impurity transport. Using conservative upper-bound
assumptions, the calculations predict rectified RF
sheath potentials of approximately 1--3~kV on the
antenna limiter sidewalls, increasing the
area-integrated gross W erosion from the antenna
limiters by a factor of approximately 64 relative to
thermal sheath conditions, yielding a gross erosion
source of $3.34\times10^{18}$~s$^{-1}$. The erosion
pattern is governed by the combined effects of
RF-modified IEADs and local incident plasma flux
rather than by RF sheath voltage alone. Ne$^{2+}$,
Ne$^{8+}$, Ne$^{3+}$, and Ne$^{4+}$ provide the
largest contributions to the primary plasma-ion
sputtering source. Approximately 10\% of the
sputtered W is locally redeposited, yielding a net
erosion source of $3.01\times10^{18}$~s$^{-1}$.
Although RF sheath acceleration substantially
enhances the localized antenna source, it remains
approximately three orders of magnitude smaller than
the total thermal divertor source and more than two
orders of magnitude smaller than the integrated
thermal main-chamber source predicted for the same
ITER plasma background. After 100~ms, approximately
22\% of the mobile W inventory remaining in the
plasma resides within the SOLPS-covered
confined-plasma region, corresponding to an annular W
concentration of
$c_{\mathrm{W,ann}}=1.70\times10^{-6}$. These
results indicate that the ITER ICRH antenna is
unlikely to dominate the overall W source budget
under the conditions considered, while demonstrating
that predictive assessment of RF-induced
plasma--material interaction requires the coupled
treatment of RF wave propagation, sheath formation,
geometry-specific IEADs, sputtering, redeposition,
and global impurity transport.
\end{abstract}
%\maketitle
\ioptwocol
\section{Introduction}

Ion cyclotron radio-frequency heating (ICRH) is a key
component of ITER's auxiliary heating system,
providing up to 10~MW of RF power for plasma heating,
supporting ion cyclotron wall conditioning (ICWC),
and enhancing operational flexibility during
high-confinement mode (H-mode)
operation~\cite{Krivska:2026,Helou2025ITERICRF}. Together with neutral beam injection and electron cyclotron resonance heating, ICRH can play an essential role in achieving and sustaining burning-plasma conditions.
\footnotetext{This manuscript has been authored by UT-Battelle, LLC, under contract DE-AC05-00OR22725 with the US Department of Energy (DOE). The US government retains and the publisher, by accepting the article for publication, acknowledges that the US government retains a nonexclusive, paid-up, irrevocable, worldwide license to publish or reproduce the published form of this manuscript, or allow others to do so, for US government purposes. DOE will provide public access to these results of federally sponsored research in accordance with the DOE Public Access Plan (http://energy.gov/downloads/doe-public-access-plan).}

Operation of high-power ICRH systems is, however, accompanied by enhanced plasma--material interactions (PMIs) at antenna structures and nearby plasma-facing components (PFCs). These interactions arise primarily from rectified RF sheath potentials generated by  parallel RF electric field ($E_\parallel$) near material surfaces. The resulting time-averaged sheath potentials can reach several hundred volts in present-day devices, for example in  WEST~\cite{Tierens_2024,kumar:2025,kumar_rfppc_2025,Tierens_2026} and are predicted to approach upto $600$~V in ITER~\cite{colas:2025}, where it is possible to have strong slow-wave coupling which may enhance sheath rectification according to the recent full-wave calculations\cite{Tierens2025_NF_ITER}. RF sheath acceleration increases the impact energy of fuel ions together with intrinsic and seeded impurities, thereby enhancing sputtering, material erosion, and impurity generation.

Following ITER's recent adoption of an all-tungsten (W) first wall and divertor~\cite{Loarte2024_ITER_Report,Loarte:2025}, understanding W erosion and impurity transport has become increasingly important. Although W provides excellent thermal performance and low tritium retention, its high atomic number makes even trace concentrations in the confined plasma capable of producing significant radiative power losses. This issue becomes particularly important during impurity-seeded operation, where Ne or Ar are introduced to enhance edge radiation and divertor power exhaust. Recent ITER modelling of neon-seeded plasmas using wide-grid SOLPS-ITER plasma backgrounds has shown that thermal plasma-wall interactions alone can generate substantial tungsten sources from both the main chamber and divertor, with the resulting impurity inventory governed by the coupled processes of erosion, redeposition, and global migration throughout the vessel~\cite{Schmid:2024, Baumann2026}. However, these studies do not include RF sheath formation or the corresponding RF-driven modification of ion acceleration at ICRH antenna structures. Under RF sheath acceleration, all Ne charge states acquire higher impact energies, with the strongest relative increase in W sputtering expected for lower charge states, while highly ionized Ne may already exhibit substantial yields under thermal-sheath conditions~\cite{colas:2025,BOBKOV:2024}.

Experimental observations in metallic-wall tokamaks further demonstrate that full behavior of the RF-induced PMI and impurity content in the plasma is affected by the coupled processes of erosion, redeposition, migration and global impurity transport in addition to the local sputtering. In WEST, which employs an ITER-like all-W PFCs environment, integrated modeling together with dedicated experiments has shown that RF sheath rectification redistributes tungsten sources from the divertor toward the ICRH antenna~\cite{Klepper:2022,colas:2023,kumar_rfppc_2025,Hakola:2026}. These studies demonstrated that the resulting impurity behavior depends on the combined effects of RF-enhanced sputtering, local incident plasma flux,  redeposition, and edge transport rather than on RF sheath voltage alone. More recently, integrated STRIPE simulations showed that variations in edge plasma density substantially modify the balance between gross erosion, redeposition, and global tungsten transport, highlighting that plasma contamination is governed by the interplay between RF sheath acceleration and scrape-off-layer transport rather than by local sputtering alone~\cite{Urbanczyk2026}.

These developments motivate the need for an integrated multi-physics framework capable of coupling edge plasma transport, RF wave propagation, sheath formation, sputtering, redeposition, and three-dimensional impurity transport for predictive assessment of RF-induced PMI in ITER. The STRIPE (Simulated Transport of RF Impurity Production and Emission) framework~\cite{kumar:2025,kumar_rfppc_2025, Kumar:2026a}  addresses this need by combining plasma backgrounds from SOLPS-ITER, full-wave RF sheath calculations in COMSOL, sputtering yields from RustBCA, charge-dependent ion energy--angle distributions computed with GITR, and three-dimensional impurity transport using GITRm. Previous applications of STRIPE to WEST~\cite{kumar:2025,kumar_rfppc_2025,Urbanczyk2026} and DIII-D\cite{Kumar:2026a} successfully reproduced main characteristics of RF sheath-driven  erosion, (re)-deposition and impurity migration, providing validation of the integrated workflow from RF sheath formation to global impurity transport.

Building upon these developments, this work
presents the first predictive application of STRIPE to
the ITER ICRH antenna. Using wide-grid SOLPS-ITER
plasma background for the ITER baseline scenario
with Ne seeding, we investigate RF sheath-driven
tungsten erosion, local redeposition, and subsequent
global impurity transport originating from the
antenna limiters. Particular emphasis is placed on the role of RF sheath acceleration of individual Ne charge states, the interplay between RF-modified ion energy--angle distributions and incident plasma flux, and the resulting transport of antenna-generated tungsten toward the confined plasma. The present work provides a more sophisticated and geometrically better resolved, predictive upper-bound modelling of the RF-specific contribution to the tungsten source and tungsten content in plasma compared to that reported previously in references\cite{colas:2025, BOBKOV:2024}. At the same time, the work establishes an integrated framework for optimizing RF antenna designs and impurity control strategies in future reactor-class fusion devices.

Together with recent work on thermal PMI studies in ITER with same plasma background\cite{Baumann2026}, the present work provides the missing RF-specific contribution to predictive tungsten source modelling in ITER and establishes an integrated framework for optimizing RF antenna designs and impurity control strategies in future reactor-class fusion devices.

This paper is organized as follows. Section~\ref{sec:3} describes the ITER edge plasma background obtained from the wide-grid SOLPS-ITER simulation together with the plasma-profile extrapolation employed to extend the solution into the recessed ICRH antenna cavity. Section~\ref{sec:4} presents the RF wave propagation, slow-wave physics, and RF sheath formation at the ITER ICRH antenna. Section~\ref{sec:5} investigates RF-driven tungsten source generation, including RF-modified ion energy--angle distributions, sputtering yields, local erosion, redeposition, and area-integrated erosion under both thermal and RF sheath conditions. Section~\ref{sec:6} presents the three-dimensional global transport of RF-generated tungsten impurities and evaluates their migration through the scrape-off layer and penetration into the confined plasma. Section~\ref{sec:7} discusses the physical mechanisms governing RF-driven PMIs, the principal modeling assumptions and limitations, and the implications for ITER and future fusion power plants. Finally, Section~\ref{sec:8} summarizes the main conclusions.

\section{ITER Plasma Background }
\label{sec:3}

\subsection{Wide-Grid SOLPS-ITER Setup with Neon Seeding}
\label{sec:3.1}

The plasma background used throughout the STRIPE workflow is obtained from the wide-grid SOLPS-ITER ($v3.2.0$) simulation corresponding to IMAS case~\#123361~\cite{Dekeyser:2021,Shtyrkhunov2024_NF}. This case represents one of a family of physically plausible ITER $Q=10$ burning-plasma scenarios and was selected because it provides comparatively unfavorable first-wall erosion conditions under thermal-sheath assumptions, where W sputtering is predominantly driven by Ne. It should therefore not be interpreted as a unique or definitive ITER baseline background, as future scenarios are expected to be further optimized, including the amount of  Ne seeding. The simulation has $P_{\rm sep}=100$~MW, equally divided between ions and electrons, and treats D as the main plasma species due to  current modeling limitation in SOLPS, although a $Q=10$ plasma is nominally expected to contain an approximately 50/50 D--T fuel mixture. Ne and He are included as seeded and intrinsic impurities, respectively. A core D fueling rate of $1\times10^{22}$~s$^{-1}$ and an additional gas puff of $5\times10^{22}$~s$^{-1}$ are prescribed, while Ne is feedback-controlled to maintain $C_{\rm Ne,sep}\simeq1.2$\%.

Unlike conventional SOLPS-ITER simulations that
terminate near the second separatrix, the wide-grid
implementation extends the computational mesh
throughout the far scrape-off layer (SOL), providing
a self-consistent plasma solution from the confined
plasma to the main-chamber plasma-facing components
(PFCs). This extended computational domain is
particularly important for RF plasma--material
interaction (PMI) studies because it resolves the
plasma immediately upstream of the ICRH antenna,
thereby eliminating the need for artificial boundary
conditions at an arbitrary radial location.

The magnetic equilibrium corresponds to the ITER
single-null baseline configuration with toroidal
magnetic field $B_t=5.3$~T and plasma current
$I_p=15$~MA. All PFCs are assumed to be W,
consistent with the latest ITER first-wall
configuration~\cite{Loarte2024_ITER_Report,Loarte:2025}.
The resulting separatrix plasma conditions are
approximately
$n_e=3.7\times10^{19}$~m$^{-3}$ and
$T_e=100$~eV.

Neutral particles are treated kinetically, while
drift terms are neglected in both the particle and
current balance equations~\cite{Shtyrkhunov2024_NF}.
Current continuity is solved together with a
self-consistent electrostatic potential, and Bohm
boundary conditions are imposed at the plasma--wall
interface to determine the sheath boundary
conditions along all PFCs. The baseline cross-field
transport coefficients are
$D_\perp=0.3$~m$^2$~s$^{-1}$ and
$\chi_{e,i}=1.0$~m$^2$~s$^{-1}$, with enhanced
transport prescribed approximately 3~cm outside the
separatrix at the outer midplane (OMP) to reproduce
profile broadening and density shoulder formation
associated with filamentary
transport~\cite{Shtyrkhunov2024_NF}. The resulting
solution represents a semi-detached divertor regime
with realistic magnetic connectivity between the
upstream SOL, divertor legs, and the ICRH antenna.

\begin{figure*}
\centering
\includegraphics[width=\linewidth]{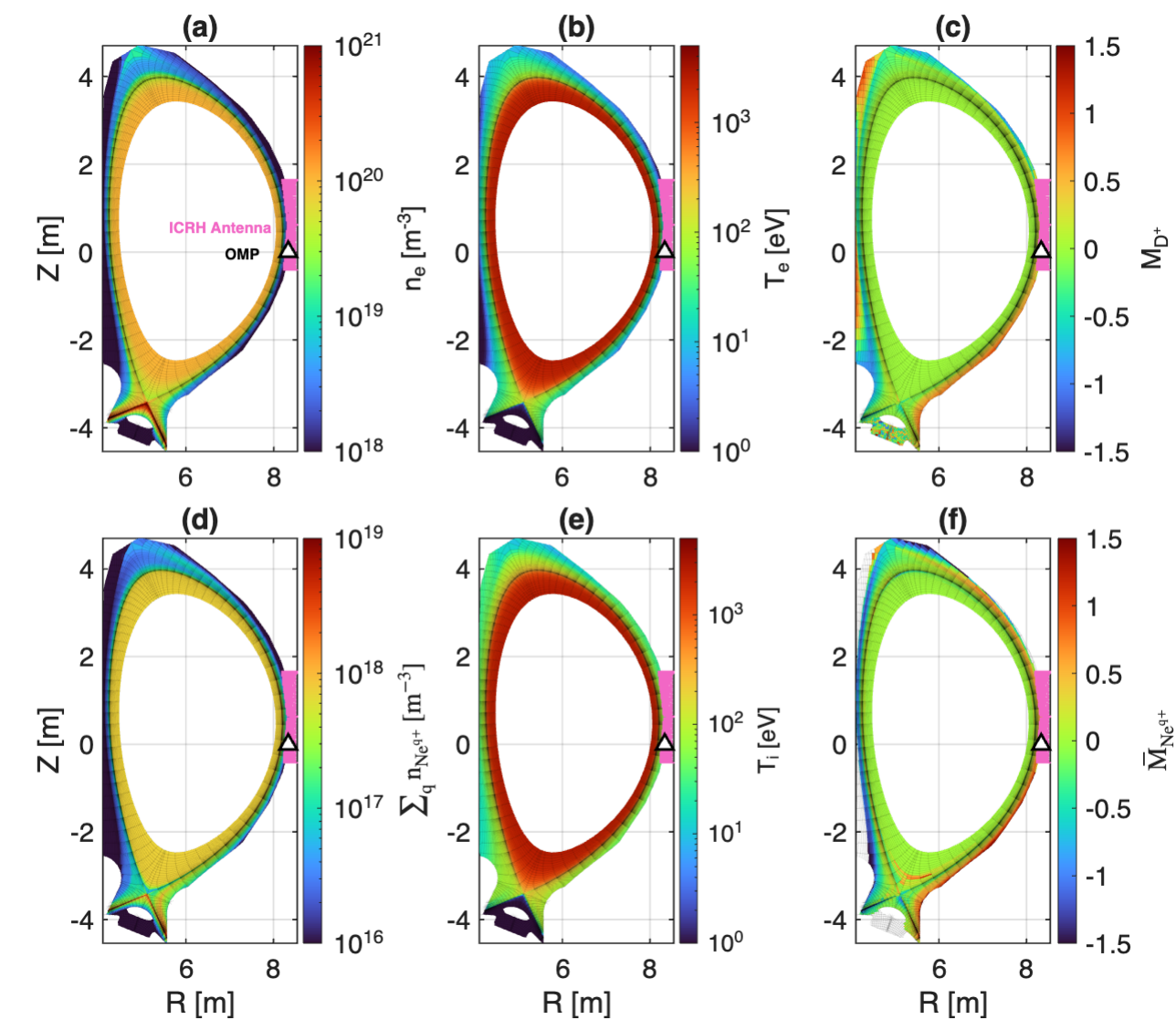}
\caption{
Two-dimensional plasma background obtained from the
wide-grid SOLPS-ITER simulation for the ITER
baseline scenario with Ne seeding. The panels show
(a) electron density ($n_e$), (b) electron
temperature ($T_e$), (c) deuterium Mach number
($M_{D^+}$), (d) total Ne ion density
($\sum_q n_{\rm Ne^{q+}}$), (e) ion temperature
($T_i$), and (f) mean Ne Mach number,
$\bar{M}({\rm Ne}^{q+})$. The pink shaded region
denotes the ITER ICRH antenna, while the triangle
marks the OMP reference location at the antenna used
throughout the one-dimensional profile analysis.
SOLPS-ITER assumes a common ion temperature for all
ion species; consequently, the deuterium and Ne
ion-temperature distributions are identical and
differ only in their corresponding density,
charge-state distributions, and flow profiles.
}
\label{fig:solps_profiles}
\end{figure*}

Figure~\ref{fig:solps_profiles} presents the
two-dimensional plasma background obtained from the
wide-grid SOLPS-ITER simulation for the ITER
baseline scenario with Ne seeding. The electron
density, temperature, Mach number, and impurity
density distributions exhibit the characteristic
features of a single-null divertor configuration,
including pronounced gradients across the
separatrix, strong poloidal asymmetry, and magnetic
connectivity between the upstream SOL, divertor, and
ICRH antenna.

Because SOLPS-ITER assumes a common ion temperature
for all ion species, the deuterium and Ne
ion-temperature distributions are identical and
differ only through their corresponding densities,
charge-state distributions, and flow fields. The electron ($T_e$) and ion ($T_i$) temperatures increase toward the separatrix within the SOL and continue to rise toward the plasma core, while decreasing along open magnetic field lines toward the plasma-facing components. The electron and Ne densities similarly decrease along open field lines toward the antenna region. The deuterium and Ne
The D and Ne Mach-number distributions indicate a relatively stagnant SOL near the antenna region. 

The ITER ICRH antenna, highlighted by the pink
shaded region in
Fig.~\ref{fig:solps_profiles}, is located on open
magnetic field lines connected directly to the
upstream SOL. The triangle marks the second separatrix reference
location used in the one-dimensional profile
analysis presented in the following sections.

\subsection{Plasma Background Extrapolation to the Antenna Region}
\label{sec:3.2}

\begin{figure}
\centering
\includegraphics[width=\linewidth]{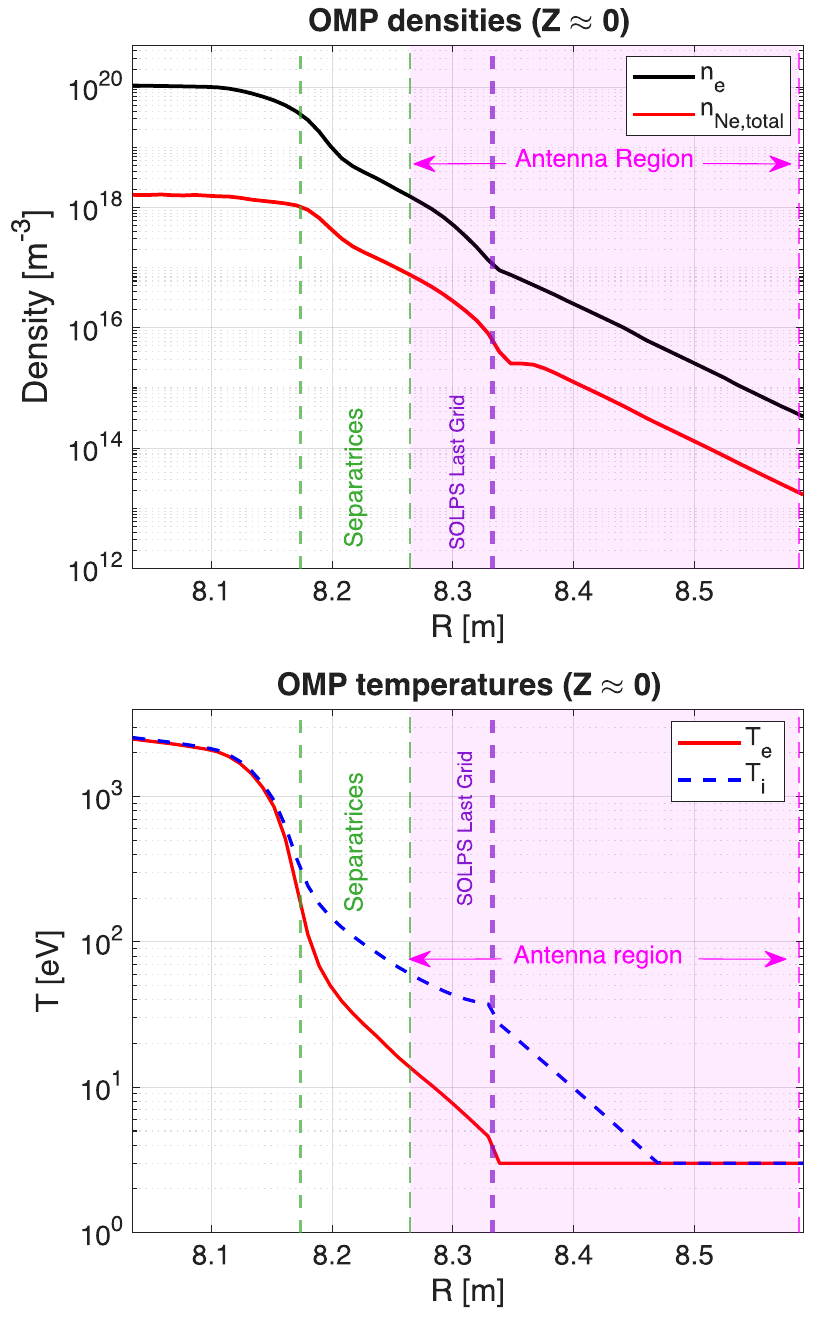}
\caption{
Radial plasma profiles extracted along the outer
midplane (OMP) passing through the ITER ICRH antenna
aperture. (a) Electron density ($n_e$) and total Ne
density ($\sum_q n_{\rm Ne^{q+}}$). (b) Electron and
ion temperatures ($T_e$ and $T_i$). The green dashed
lines denote the separatrices, while the purple
vertical line indicates the last valid SOLPS-ITER
grid point at the OMP. Beyond this location, the
plasma background is extrapolated into the recessed
antenna cavity. The pink shaded region denotes the
extent of the ITER ICRH antenna.
}
\label{fig:omp_profiles}
\end{figure}

Although the wide-grid SOLPS-ITER simulation extends
substantially farther into the far SOL than
conventional edge-plasma simulations, the
computational mesh does not fully resolve the
recessed ITER ICRH antenna box and the Faraday-screen
(FS) bars where RF sheath formation occurs.
Consequently, the plasma quantities required by the
STRIPE workflow are extrapolated beyond the SOLPS
computational boundary to provide continuous plasma
conditions throughout the antenna cavity and on the
FS surfaces.

Figure~\ref{fig:omp_profiles} presents the radial
plasma profiles extracted along the OMP passing
through the antenna aperture. The green dashed lines
indicate the two separatrices, while the purple
vertical line marks the last valid SOLPS-ITER grid
point, beyond which the extrapolated plasma solution
begins. The pink shaded region denotes the radial
extent of the  ITER ICRH antenna (see
Fig.~\ref{fig:app_density}(b) for the corresponding
radial-coordinate map on the antenna limiter
surfaces). Across the SOL, the electron density
decreases substantially toward the antenna region,
while the electron and ion temperatures decrease
from several keV in the confined plasma to only a
few electronvolts inside the recessed antenna
cavity. These profiles provide the reference plasma
conditions used to extend the SOLPS-ITER solution
into the antenna geometry.

The extrapolation follows the methodology previously
developed for STRIPE simulations of the DIII-D
helicon antenna~\cite{Kumar:2026a}. Electron density
($n_e$), deuterium density ($n_D$), electron
temperature ($T_e$), and ion temperature ($T_i$) are
represented by least-squares exponential fits
obtained from linear regressions of
$\log(n_e)$, $\log(n_D)$,
$\log(T_e)$, and $\log(T_i)$
along the OMP. The fitted profiles are matched to the
original SOLPS solution at the computational boundary
to ensure a continuous transition while preserving
the local plasma values. Because the extrapolated
profiles are obtained from least-squares exponential
fits rather than from an additional plasma transport
calculation, small changes in the local slope may
occur near the transition region. Beyond the SOLPS
domain, the plasma parameters follow the
corresponding exponential decay as a function of the
shortest distance to the nearest valid SOLPS cell,
providing a continuous extension of the plasma
background into the recessed antenna cavity.

To avoid unphysically low temperatures in the
extrapolated region, lower bounds of
$T_e=T_i=3$~eV are imposed. These values represent
reasonable estimates for the cold edge plasma
immediately in front of the antenna and affect only
the extrapolated region outside the SOLPS
computational domain.

The adopted extrapolation is expected to provide a
conservative upper bound for the plasma density
within the recessed antenna cavity. In reality, the
effective magnetic connection length decreases
progressively inside the antenna box, leading to a
reduction of the local density e-folding length that
is not captured by the simple exponential extension.
Consequently, the present approach is expected to
overestimate the plasma density, and therefore the
incident ion flux and RF-driven sputtered W source,
within the recessed antenna region. The extrapolated
background should therefore be interpreted as
providing an upper-bound estimate for the subsequent
RF sheath-driven sputtering calculations.

The parallel flow velocities of all plasma species
are held constant outside the SOLPS domain by
assigning the values from the nearest valid SOLPS
cell, thereby avoiding artificial flow gradients.
Similarly, the relative populations of the
individual Ne charge states are held fixed at their
boundary values while the total Ne density continues
to follow the same exponential decay as the
background plasma. This approach preserves the
SOLPS-predicted local charge-state composition and
corresponding effective charge ($Z_{\rm eff}$)
throughout the extrapolated antenna region without
introducing additional assumptions regarding
nonequilibrium ionization outside the SOLPS domain.

The extrapolated plasma fields are subsequently
mapped onto the three-dimensional COMSOL and
GITR/GITRm meshes assuming toroidal symmetry of the
background plasma. The extrapolation serves only to
extend the SOLPS plasma solution into the antenna
region and does not constitute an additional plasma
transport calculation.

\subsection{Impurity Charge-State Distribution in the Antenna Region}
\label{sec:3.3}

\begin{figure}
\centering
\includegraphics[width=\linewidth]{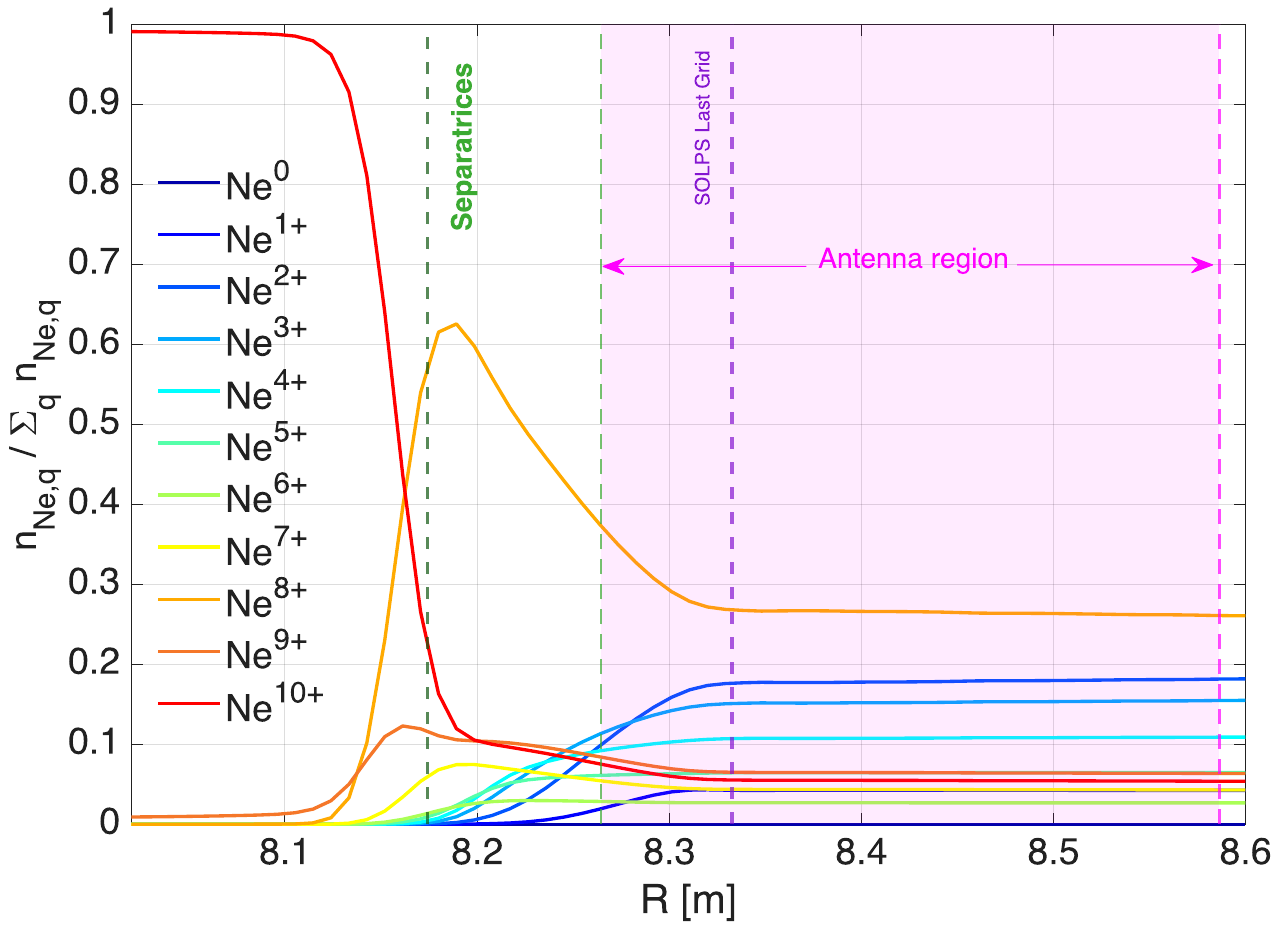}
\caption{
Radial evolution of the Ne charge-state fractions
along the outer midplane (OMP). The separatrix,
antenna aperture, and recessed antenna region are
indicated. Within the SOLPS computational domain, the
charge-state fractions evolve self-consistently with
the local plasma conditions. Beyond the SOLPS
boundary, the relative Ne charge-state fractions are
held fixed at their boundary values during the plasma
extrapolation, while the total Ne density continues
to decrease exponentially. Fully stripped
Ne$^{10+}$ dominates near the separatrix, whereas
Ne$^{8+}$ becomes the dominant charge state
throughout the antenna region.
}
\label{fig:ne_charge}
\end{figure}

The impurity charge-state distribution strongly
influences RF-induced sputtering because the energy
gained across the RF sheath scales approximately with
the ion charge state. Figure~\ref{fig:ne_charge}
presents the evolution of the Ne charge-state
fractions along the OMP from the confined plasma into
the recessed antenna region. Within the SOLPS
computational domain, the charge-state populations
evolve self-consistently in response to the local
plasma density and temperature. Beyond the SOLPS
computational boundary, the relative charge-state
fractions remain fixed at their boundary values, as
described in Section~\ref{sec:3.2}, while only the
total Ne density continues to decrease
exponentially. 

Close to the separatrix, where the electron
temperature exceeds 100~eV, the
impurity population is nearly fully ionized and is
dominated by Ne$^{10+}$. As the plasma expands into
the SOL and cools toward the antenna, the charge-state
distribution progressively shifts toward lower
ionization states.

Within the antenna region, Ne$^{8+}$ becomes the
dominant charge state, while lower charge states,
particularly Ne$^{2+}$--Ne$^{4+}$, remain a
significant fraction of the total Ne population.
Although these lower charge states experience less RF
sheath acceleration than Ne$^{8+}$ or Ne$^{10+}$,
their comparatively large abundance allows them to
contribute appreciably to the impurity-driven
sputtering source. By contrast, Ne$^{1+}$ and
neutral Ne constitute only a small fraction of the
modeled impurity population within the antenna
region.

The intrinsic He concentration is comparatively low
and remains almost entirely in the fully ionized
He$^{2+}$ state throughout the SOL. Consequently, He
contributes only weakly to the overall impurity
composition compared with the seeded Ne population
and is therefore neglected in the subsequent erosion
and transport analysis.

\section{RF Wave Propagation and Sheath Formation}
\label{sec:4}

The rectified RF sheath potentials developing on the ITER ICRH antenna limiters are computed using the sheath-equivalent dielectric-layer technique~\cite{Beers1,Beers2,Tierens_2024}. In this approach, thin dielectric layers of thickness $d=2$~mm are applied to the limiter surfaces to represent the RF sheath through spatially varying effective permittivity ($\varepsilon_{\rm eff}$) and conductivity ($\sigma_{\rm eff}$). These effective material properties are obtained from the nonlinear sheath impedance model\cite{myra2017physics}, which depends on the local plasma density, temperature, magnetic-field orientation, and RF potential across the sheath layer. The artificial dielectric satisfies
$\lambda_D \ll d \ll \lambda_{\rm RF}$,
thereby embedding the RF sheath response directly into the electromagnetic material properties without requiring explicit sheath boundary conditions, as employed in alternative formulations such as Petra-M~\cite{shiraiwa2023magnetic}, STIX~\cite{Migliore_2023}, and related full-wave RF sheath models.

The RF calculations employ the plasma background
described in Section~\ref{sec:3}, consisting of the
wide-grid SOLPS-ITER solution together with its
extrapolated extension into the recessed antenna
cavity. This plasma background provides continuous
distributions of $n_e$, $T_e$, and $T_i$ throughout
the antenna region, enabling physically consistent RF
sheath calculations on the antenna limiter
structures. In the present work, the RF sheath
solution is computed only for these limiter surfaces,
which constitute the RF-driven W source considered in
the subsequent erosion calculations. Other
magnetically or electrically connected plasma-facing
components are not included in the COMSOL RF sheath
model. By contrast, the Petra-M reference simulations
summarized in ~\ref{app:rfsheath} treat the
entire antenna panel, thereby providing additional
qualitative information on RF-active regions outside
the limiter surfaces considered in the present
erosion model.

For comparison,
~\ref{app:rfsheath} summarizes two reference
ITER edge-plasma scenarios derived from the 2010 ITER
baseline equilibrium~\cite{ITER_D_33Y59M},
representing low- and high-density operating
conditions for RF sheath formation. Compared with
these reference profiles, the present wide-grid
SOLPS-ITER solution provides a self-consistent plasma
background in which portions of the antenna reside in
both the slow-wave-propagative ($S>0$) and
slow-wave-evanescent ($S<0$) regimes. In the
immediate vicinity of the recessed antenna cavity,
the local plasma density is lower than that of the
2010 low-density reference case. However, higher
far-SOL densities may be expected for scenarios with
reduced Ne seeding and, particularly, for wide-grid
backgrounds including a sub-divertor description,
where recently published SOLPS-ITER calculations show
an increase in far-SOL density of more than an order
of magnitude~\cite{Pshevov:2025}. Furthermore, the
present simulations do not include localized gas
puffing in front of the ICRH antenna, which may be
employed during ITER operation to increase the local
edge density and improve RF
coupling~\cite{Zhang_2017,Zhang_2023}. These density
modifications could alter the local slow-wave
accessibility and resulting RF sheath structure. A
quantitative assessment of these effects is beyond
the scope of the present work. The corresponding
COMSOL and Petra-M calculations presented in
~\ref{app:rfsheath} illustrate the
qualitative sensitivity of RF sheath formation to
edge plasma density and slow-wave accessibility.
While Petra-M predicts RF-active regions over the
full antenna panel, the present erosion calculations
are restricted to the limiter surfaces for which the
self-consistent COMSOL RF sheath solution is
available.

Consistent with the ITER baseline full-field operating
scenario~\cite{Krivska:2026,Helou2025ITERICRF}, a
case with 7~MW of coupled RF power is considered.
The 24 antenna feed ports are excited in a balanced
configuration, with ports 1--3 and 13--15 phased at
$-2.9$ rad, ports 4--6 and 16--18 at
$1.7$ rad, ports 7--9 and 19--21 at
$0.2$ rad, and ports 10--12 and 22--24
at $-1.4$ rad. This case is a good representative for the upper-bound RF-specific tungsten source calculations, because the chosen antenna feeding is not optimized to reduce the RF sheaths. Under these plasma conditions, portions of the antenna reside in the slow-wave propagative regime, allowing localized slow-wave coupling and resonance-cone formation that enhance RF sheath rectification.

A four-step iterative procedure is employed to obtain self-consistent RF sheath solutions:

\begin{enumerate}
\item Perform an initial full-wave calculation assuming vacuum sheath conditions ($\varepsilon_{\rm eff}=1$, $\sigma_{\rm eff}=0$).
\item Compute the local sheath impedance $Z_{\rm sh}$ and determine updated dielectric properties from
\begin{equation}
\sigma_{\rm eff}
+i\omega\epsilon_0\epsilon_{\rm eff}
=
\frac{d}{Z_{\rm sh}},
\label{eq:sheath_layer}
\end{equation}
where $\epsilon_{\rm eff}$ is the relative effective
permittivity of the sheath-equivalent dielectric layer,
$\sigma_{\rm eff}$ is its effective conductivity,
$\epsilon_0$ is the vacuum permittivity and $d$ is the sheath layer thickness.
\item Update the dielectric-layer properties in COMSOL.
\item Repeat until the sheath impedance converges.
\end{enumerate}

Convergence is typically achieved after five iterations, producing stable spatial distributions of both the sheath impedance and the rectified RF sheath potential. The converged solution therefore represents a nonlinear equilibrium between the RF wave field and the sheath impedance within the cold-plasma approximation.

\begin{figure}
    \centering
    \includegraphics[width=\linewidth]{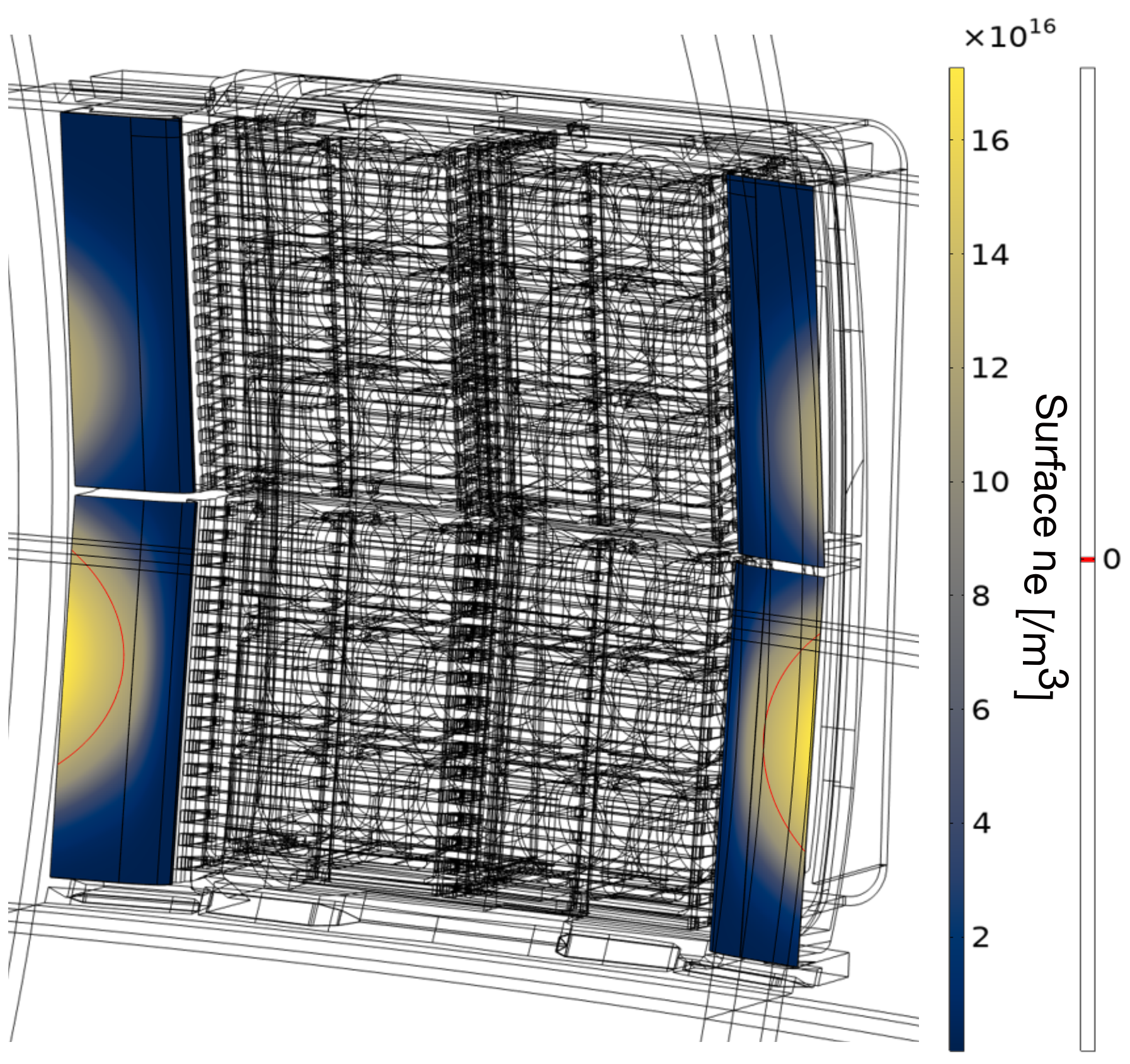}
    \caption{Spatial distribution of the electron density near the ITER ICRH antenna obtained from the wide-grid SOLPS-ITER plasma background together with contours of the cold-plasma Stix parameter $S=0$ (red). The $S=0$ contour separates slow-wave-accessible ($S>0$) and slow-wave-evanescent ($S<0$) regions. Under the present ITER baseline conditions, the recessed limiter sidewalls adjacent to the Faraday-screen bars remain within the $S>0$ region, allowing slow-wave propagation and resonance-cone formation immediately in front of the antenna.}
    \label{fig:iter_antenna_ne}
\end{figure}

Figure~\ref{fig:iter_antenna_ne} shows the electron-density distribution surrounding the ITER ICRH antenna together with contours of the cold-plasma Stix parameter $S=0$. The $S=0$ contour separates slow-wave-accessible ($S>0$) and slow-wave-evanescent ($S<0$) regions. Under the chosen ITER baseline conditions, the limiter sidewalls adjacent to the Faraday-screen bars remain predominantly within the $S>0$ region, whereas only portions of the front limiter surfaces intersect the $S<0$ region. Consequently, the slow wave propagates directly along the recessed sidewalls, allowing resonance-cone formation and localized enhancement of the parallel RF electric field. 

\begin{figure}
    \centering
    \includegraphics[width=\linewidth]{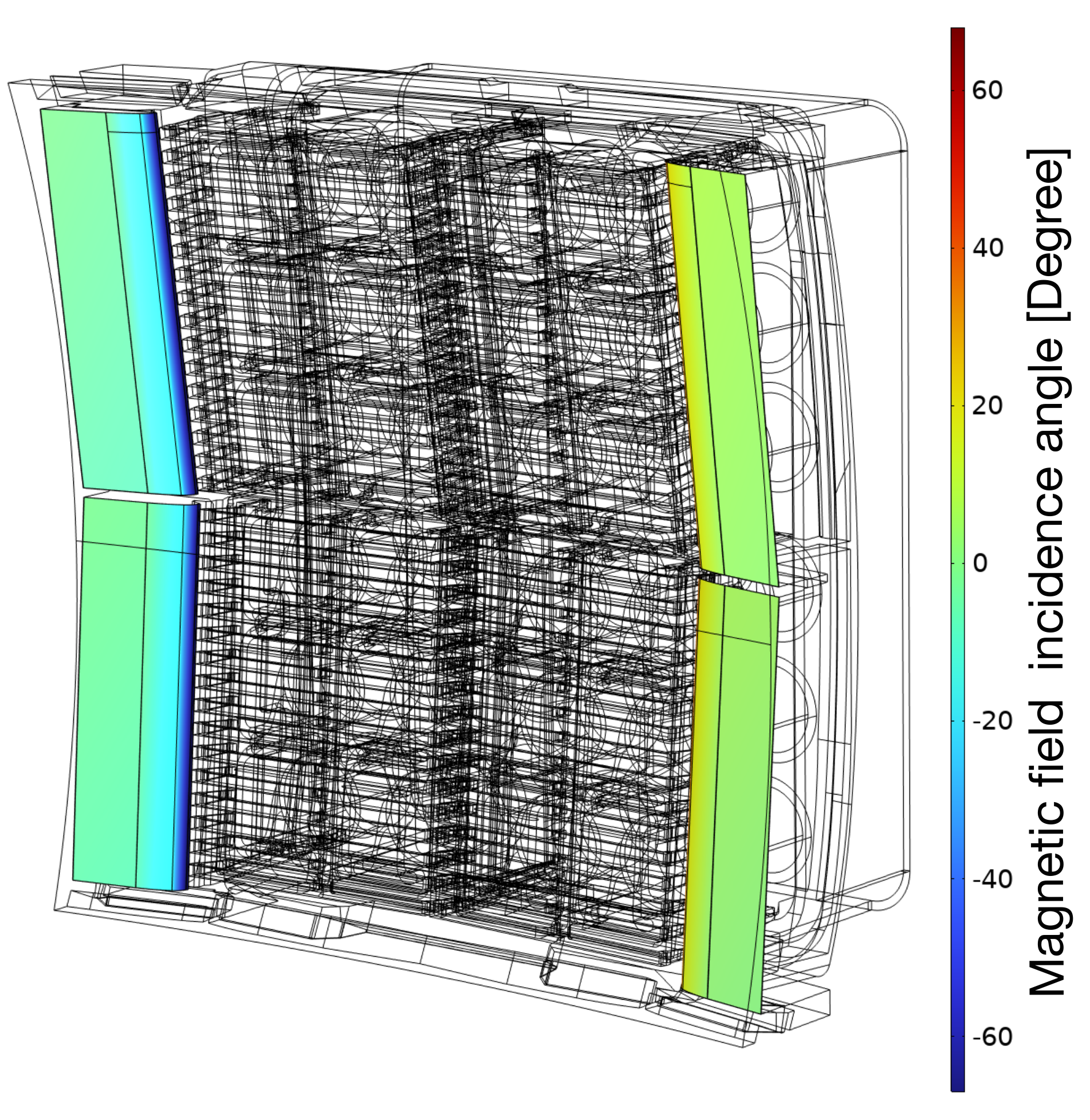}
    \caption{Angle between the magnetic field and the ITER ICRH limiter surfaces. Owing to the curved limiter geometry, the magnetic field becomes increasingly grazing toward the recessed antenna sidewalls, enhancing the importance of the magnetic pre-sheath in the Brooks sheath formulation.}
    \label{fig:Bfield_angle}
\end{figure}

The effectiveness of RF sheath formation also depends strongly on the magnetic-field incidence angle relative to the PFCs. Figure~\ref{fig:Bfield_angle} presents the magnetic-field incidence angle over the active limiter surfaces. The magnetic field is nearly tangential to the limiter faces over most of the antenna, with only modest spatial variation arising from the curved limiter geometry. The opposite sign of the incidence angle on the left and right limiters reflects the change in surface normal rather than a change in magnetic-field direction. Within the Brooks sheath formulation, the local magnetic-field incidence angle determines the partitioning of the total sheath potential between the Debye sheath and the magnetic pre-sheath (Chodura sheath), thereby modifying the local sheath electric field and the trajectories of sheath-accelerated ions. Consequently,  the strongest RF sheath effects in the calculations are expected where favorable magnetic-field incidence coincides with slow-wave-accessible ($S>0$) regions, rather than being determined by magnetic geometry alone.  

Although the sheath-equivalent dielectric-layer formulation provides a self-consistent description of RF sheath formation within the cold-plasma approximation, the predicted sheath voltages should be interpreted with appropriate caution. Recent validation studies on WEST~\cite{Tierens_2024,kumar:2025,kumar_rfppc_2025,Tierens_2026} have shown that cold-plasma RF models accurately reproduce the observed spatial localization of RF sheath formation but generally overpredict the absolute rectified sheath voltages, particularly under low-density conditions where the slow wave becomes propagative and resonance-cone structures dominate the RF electric field, as confirmed by recent emissive-probe measurements on WEST~\cite{Diab:2025, Diab:2026}. The experiments in low-density conditions on ASDEX Upgrade also do not show significantly increased RF currents on the antenna limiters and W erosion compared to the conditions when the slow wave is evanescent\cite{BOBKOV:2024}. These discrepancies arise from the combined effects of the cold-plasma approximation, the singular behavior of the dielectric response near the lower-hybrid resonance ($S=0$), the finite spatial resolution required to resolve the narrow slow-wave resonance, and the absence of kinetic wave–particle interactions that naturally regularize the RF fields. In addition, other non-linear effects that reduce the rectified sheath voltages, such as the DC current transport, are not taken into account and the antenna feeding is not optimized to reduce the sheath voltages. Consequently, for the present calculations, the absolute sheath amplitudes should be regarded as conservative upper-bound estimates rather than quantitative predictions. The RF sheath solution adopted throughout the present
work therefore represents the mixed slow-wave
accessibility obtained from the wide-grid SOLPS-ITER
background. The high- and low-density cases summarized
in \ref{app:rfsheath} are included only to illustrate the
sensitivity of sheath localization to the edge plasma
conditions.

\begin{figure}
    \centering
    \includegraphics[width=\linewidth]{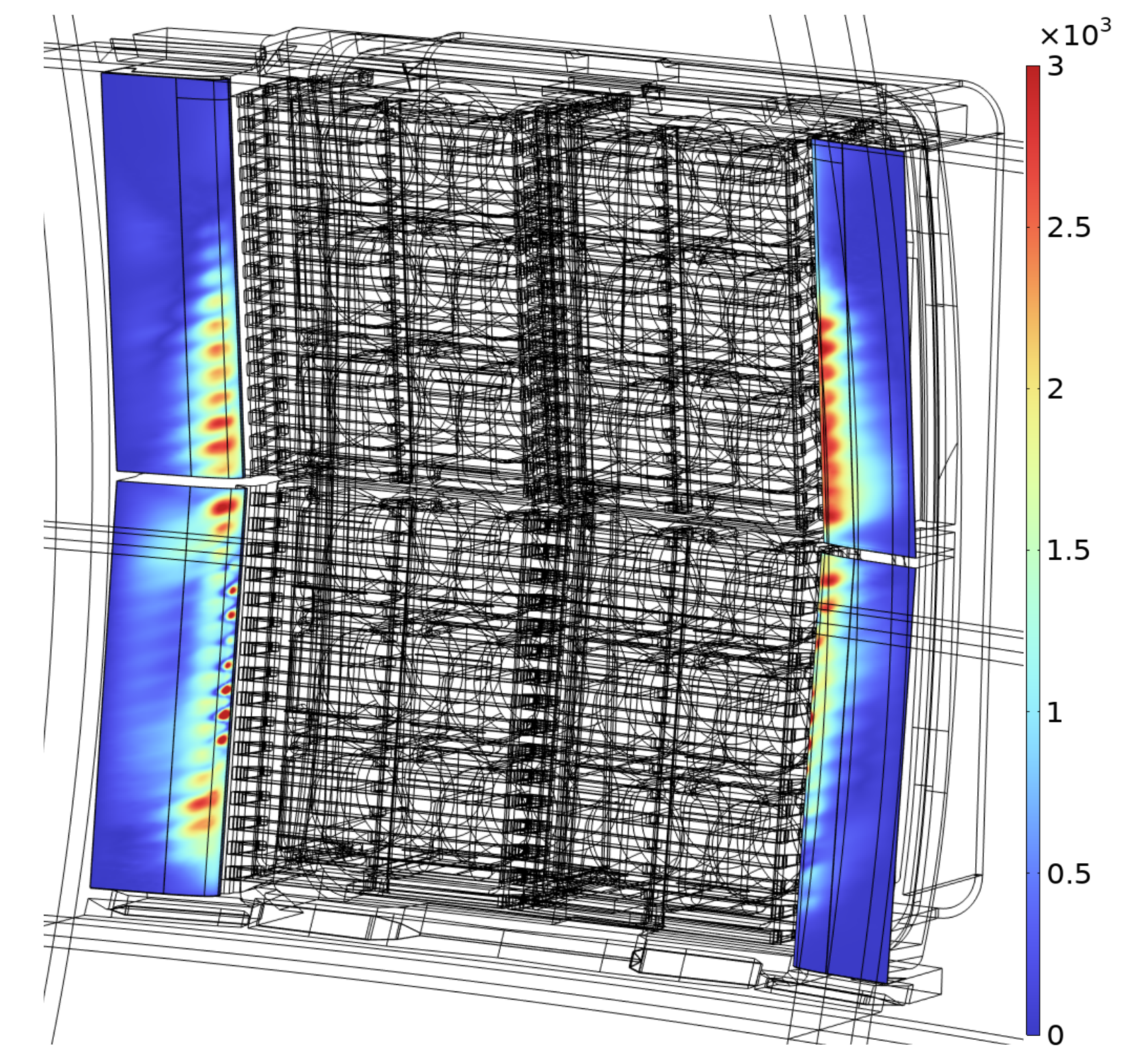}
    \caption{Rectified DC sheath potentials (\(V_{\mathrm{RF,DC}}\)) on the ITER ICRH antenna limiters from COMSOL simulations using the wide-grid \textit{SOLPS-ITER} background (IMAS~\#123361). Peak potentials up to 3~kV are localized on the sidewalls adjacent to the FS bars.}
    \label{fig:VDC_ITER}
\end{figure}

Figure~\ref{fig:VDC_ITER} presents the converged rectified RF sheath potentials obtained from the COMSOL calculations. Peak sheath voltages approaching 3~kV develop predominantly along the recessed limiter sidewalls adjacent to the Faraday-screen bars, whereas substantially smaller sheath potentials occur on the front limiter faces. The RF sheath distribution follows the combined influence of plasma density, slow-wave accessibility, and magnetic-field geometry discussed above. Regions located within the propagative slow-wave regime ($S>0$) experience enhanced parallel RF electric fields associated with resonance-cone formation, while the increasingly grazing magnetic-field incidence on the recessed sidewalls further enhances the magnetic pre-sheath contribution within the Brooks sheath model. Together,  these effects in the calculations, produce the strongest RF sheath rectification on the sidewalls rather than on the front limiter surfaces.

The predicted RF sheath distribution also exhibits a moderate top--bottom asymmetry arising from the corresponding asymmetry of the local plasma density. Because the lower limiter is exposed to a denser plasma than the upper limiter, the position of the $S=0$ contour and the extent of the slow-wave-accessible region differ slightly between the two sides of the antenna. 

It is important to note that the regions of maximum RF sheath potential do not necessarily correspond to the regions of maximum material erosion. The RF sheath determines the ion acceleration and therefore the local ion energy-angle distributions (IEADs), whereas the gross erosion additionally depends on the incident plasma flux reaching the surface. Consequently, the RF sheath map should be interpreted primarily as the distribution of available acceleration potential rather than as a direct indicator of the resulting erosion. The relationship between RF sheath acceleration, local IEADs, incident plasma flux, and the resulting tungsten source is investigated in the following section.

The converged RF sheath potentials are subsequently imported into GITR/GITRm for erosion and global impurity transport calculations.

\section{RF-Driven Tungsten Erosion}
\label{sec:5}

\subsection{Geometry Definition and Surface Selection }
\label{sec:5.1}

To couple the COMSOL sheath modeling with impurity transport in STRIPE, the ITER ICRH antenna geometry is imported into the GITR and GITRm domains. Direct use of the complete engineering CAD model is impractical for Monte Carlo particle tracking because of its numerous fine engineering details such as bolt holes, fillets, and small mechanical features. Therefore, a \textit{defeatured} geometry is generated from the original CAD model, in which all PFCs are preserved with high fidelity, while non-relevant features are removed to ensure mesh quality and computational efficiency.

\begin{figure}
    \centering
    \includegraphics[width=\linewidth]{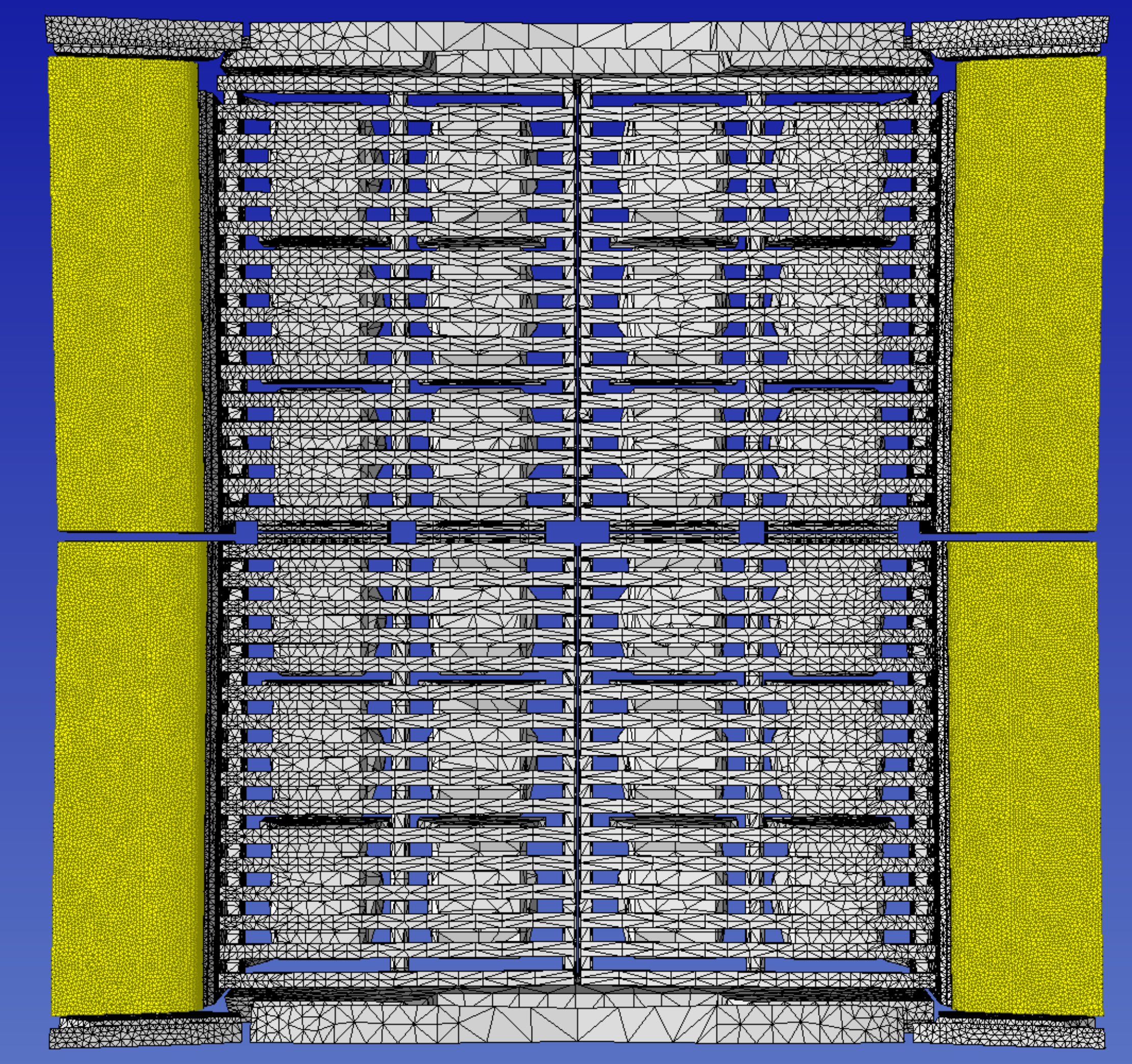}
    \caption{Defeatured geometry used in GITR/GITRm simulations showing the ITER ICRH antenna region. The limiter surfaces ({yellow}) are treated as active sheath regions where rectified potentials \(V_{\mathrm{RF,DC}}\) are applied and W erosion sources are initialized. The remaining surrounding structures ({gray}) are included as passive boundaries for evaluating impurity re-deposition and migration patterns.}
    \label{fig:geometry_GITR}
\end{figure}

In this defeatured representation, the limiter structures are defined as the \textit{active sheath surfaces}, as shown in Fig.~\ref{fig:geometry_GITR}. These limiter surfaces, highlighted in {yellow}, represent the regions where rectified potentials \(V_{\mathrm{RF,DC}}\) are applied and W erosion sources are initialized in STRIPE. The remaining surrounding structures--including the Faraday screen and other antenna components--are retained in the geometry (shown in {gray}) to capture the re-deposition of sputtered impurities and secondary transport patterns, but they are not assigned sheath voltages.

This defeatured geometry preserves the essential physics of the antenna–plasma interface while remaining computationally tractable for large-scale kinetic modeling. By treating only the limiters as active erosion sources and retaining other surfaces as passive re-deposition boundaries, GITR and GITRm efficiently resolve both local sputtering and global W migration in the ITER ICRH environment.

\subsection{Sputtering Yield Analysis for ITER-Relevant Projectiles Using RustBCA}
\label{sec:5.2}

Energy- and angle-dependent sputtering yields,
$Y(\mathcal{E},\theta)$,
are pre-computed using RustBCA for D, Ne, He, and W projectiles incident on W targets. The calculations span projectile energies from 1~eV to 10~keV for D, He, and W, while the Ne calculations are extended to 80~keV to encompass the upper energy range potentially accessible to highly charged RF-accelerated Ne ions. This extended energy range provides a conservative upper bound for evaluating RF-induced sputtering and does not imply that the majority of impurity ions reach these energies during ITER operation.

Within STRIPE, RustBCA assumes neutral projectiles, consistent with the binary-collision approximation (BCA). Charge-state effects are incorporated separately through the GITR-computed charge-dependent ion trajectories and RF sheath acceleration, which determine the IEADs. These IEADs are subsequently convolved with the neutral-projectile sputtering yields computed by RustBCA to determine the effective sputtering yields used throughout the erosion calculations.

\begin{figure*}
\centering
\includegraphics[width=\linewidth]{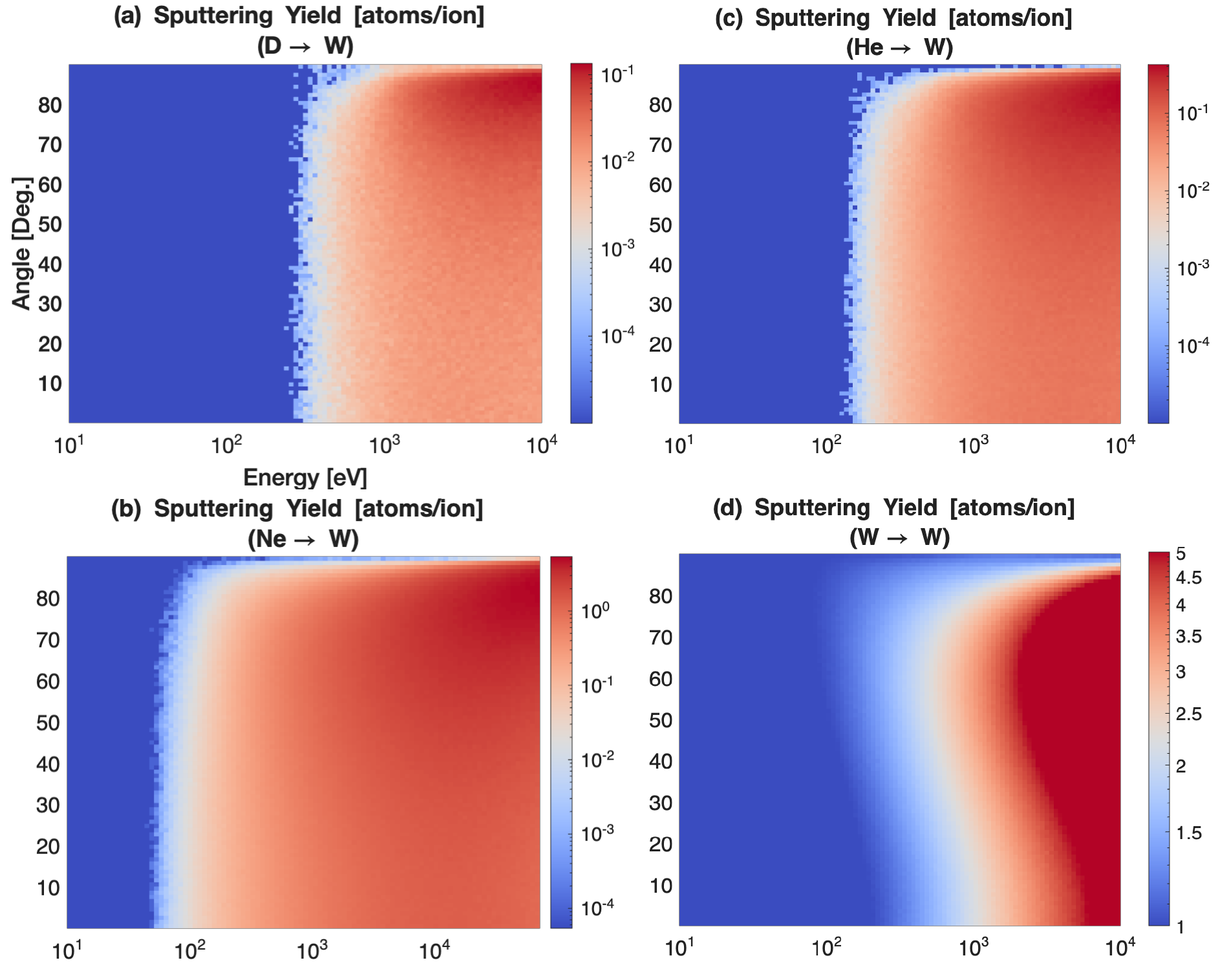}
\caption{
Two-dimensional sputtering-yield surfaces,
$Y(\mathcal{E},\theta)$,
computed using RustBCA for (a) D$\rightarrow$W, (b) Ne$\rightarrow$W, (c) He$\rightarrow$W, and (d) W$\rightarrow$W self-sputtering. The surfaces summarize the combined energy and angle dependence used throughout the effective sputtering-yield calculations.
}
\label{fig:iter_spYield_2D}
\end{figure*}

Figure~\ref{fig:iter_spYield_2D} summarizes the two-dimensional sputtering-yield surfaces,
$Y(\mathcal{E},\theta)$,
for D$\rightarrow$W, Ne$\rightarrow$W, He$\rightarrow$W, and W$\rightarrow$W self-sputtering. The yield depends strongly on both the incident energy and impact angle, with the largest sputtering yields occurring for heavy projectiles at multi-keV energies and grazing incidence. In particular, Ne and W exhibit substantially larger sputtering yields than D and He, indicating that impurity-driven sputtering is expected to dominate once energetic impurity ions are accelerated across the RF sheath.

\begin{figure*}
\centering
\includegraphics[width=\linewidth]{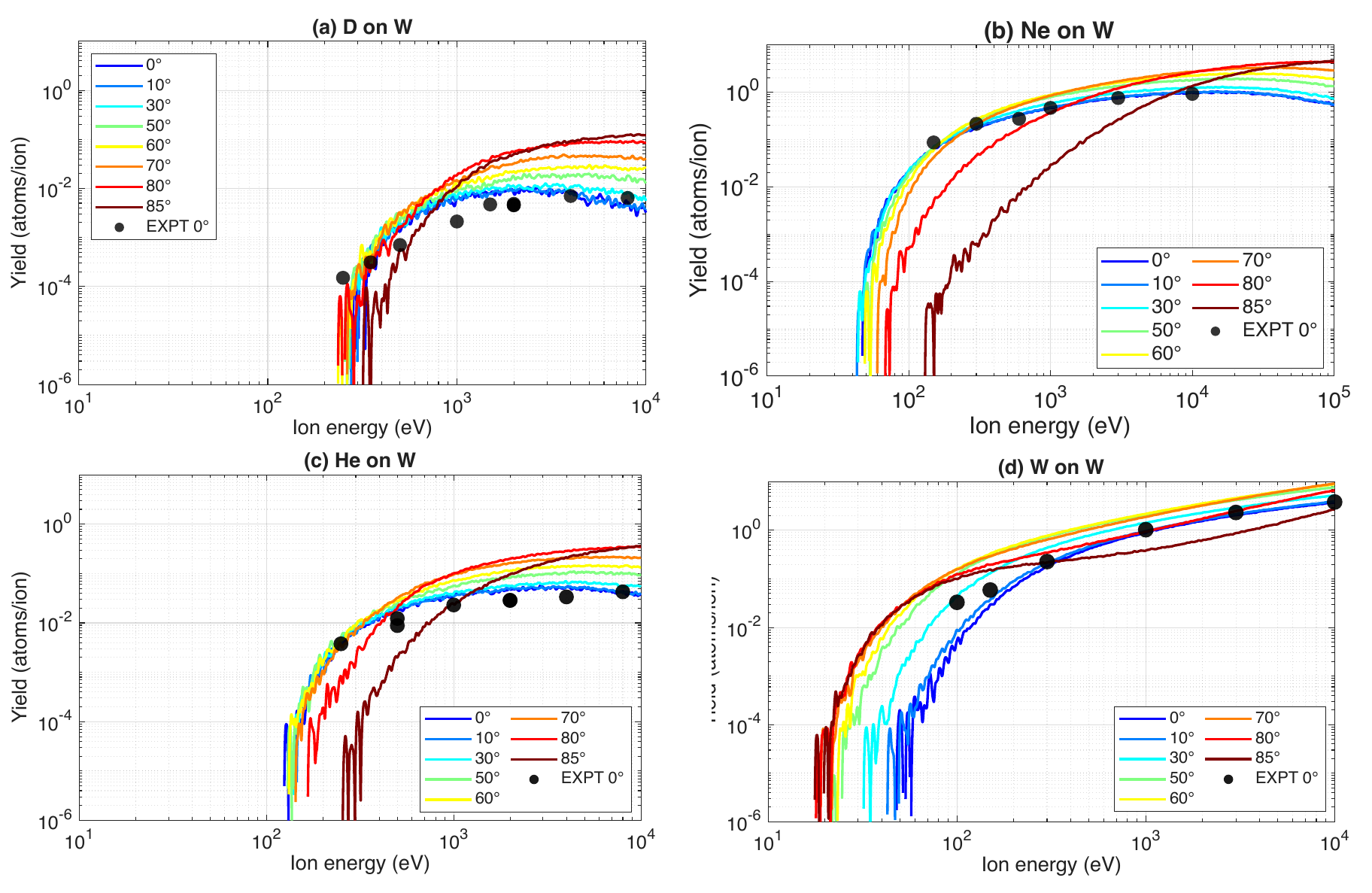}
\caption{
Energy-dependent RustBCA sputtering yields for (a) D$\rightarrow$W,
(b) Ne$\rightarrow$W, (c) He$\rightarrow$W, and
(d) W$\rightarrow$W. Curves correspond to incidence angles between
$0^\circ$ and $85^\circ$. Experimental measurements at normal incidence
($0^\circ$) from the Eckstein IPP Report~9/82 are shown as black circles
for reference. RustBCA reproduces the available experimental trends
while predicting the expected enhancement in sputtering yield with
increasing incidence angle.
}
\label{fig:rustbca_validation}
\end{figure*}

Figure~\ref{fig:rustbca_validation} presents the energy dependence of the RustBCA sputtering yields for (a) D$\rightarrow$W, (b) Ne$\rightarrow$W, (c) He$\rightarrow$W, and (d) W$\rightarrow$W. Curves are shown for incidence angles between $0^\circ$ and $85^\circ$, while the available experimental measurements at normal incidence ($0^\circ$) from the Eckstein {\it{et. al.}}, IPP Report~9/82~\cite{Eckstein1993SputteringData} are included for reference. Overall, RustBCA reproduces the available experimental trends at normal incidence while predicting the expected enhancement in sputtering yield with increasing impact angle.

As shown in Fig.~\ref{fig:rustbca_validation}(a), sputtering by D begins near a threshold energy of approximately 250--300~eV and increases gradually with projectile energy. Even at several keV, the sputtering yield remains below approximately $10^{-1}$~atoms/ion, with the highest yields occurring for grazing incidence ($\theta\gtrsim70^\circ$). Beyond a few keV, the sputtering yield saturates and subsequently decreases slightly at multi-keV energies because of the declining nuclear stopping efficiency.

In contrast, Fig.~\ref{fig:rustbca_validation}(b) shows that Ne exhibits a substantially lower sputtering threshold (approximately 40~eV) together with a much steeper increase in sputtering yield. The yield exceeds unity near 1~keV and continues to increase with projectile energy before gradually saturating beyond several keV. At multi-keV energies, the sputtering yield decreases slightly owing to the reduced nuclear stopping efficiency at higher energies. 

Figure~\ref{fig:rustbca_validation}(c) illustrates the intermediate sputtering behavior of He. The sputtering threshold is substantially lower than that of D but higher than that of Ne, and the yield increases steadily with projectile energy before gradually saturating beyond several keV. At multi-keV energies, the yield decreases slightly owing to the reduced nuclear stopping efficiency. Although He sputters W more efficiently than D, its yield remains significantly below that of Ne over the energy range considered. Furthermore, the He concentration in the present ITER scenario is very low. Consequently, He-induced sputtering is expected to make a negligible contribution to the overall tungsten erosion and impurity transport and is therefore neglected in the subsequent erosion and transport calculations.

The W self-sputtering yield in
Fig.~\ref{fig:rustbca_validation}(d) increases rapidly
above a few hundred electronvolts and continues to
rise throughout the simulated energy range up to
10~keV, exhibiting no indication of saturation. At
multi-keV energies and grazing incidence, the yield
approaches values comparable to those for
Ne-induced sputtering. Consequently, re-ionized W
accelerated by the RF sheath can provide a secondary
amplification mechanism for localized erosion once a
sufficient W population develops near the antenna.
However, a self-sustaining sputtering cascade
requires not only sputtering yields exceeding unity
but also efficient prompt re-ionization and
redeposition of sputtered W onto RF sheath-active
surfaces. Under the ITER conditions considered here,
W self-sputtering originating from the ICRH antenna remains a secondary contribution
and is not expected to produce a self-sustaining erosion
cascade.

These angle-resolved sputtering-yield databases are subsequently convolved with the GITR-computed, charge-state-dependent IEADs to determine the geometry-specific effective sputtering yields used throughout the tungsten erosion calculations. Uncertainties associated with the BCA approximation and the limited experimental database at oblique incidence are discussed further in Section~\ref{sec:7}.

\subsection{Geometry-Specific Effective Sputtering Yields}
\label{sec:5.3}

The sputtering yields computed with RustBCA represent
monoenergetic, angle-resolved impacts on an ideal
planar surface. In the ITER ICRH antenna environment,
however, ions experience sheath acceleration,
gyromotion, Coulomb collisions, plasma flow, and
magnetic-field-guided transport before reaching the
limiter surfaces. Consequently, the IEADs incident on the
material surface differ substantially from the
monoenergetic distributions used to construct the
RustBCA sputtering-yield database.

To account for these effects, GITR computes
charge-state-dependent IEADs by following ions from
the sheath edge to the plasma-facing components under
the combined influence of the magnetic field,
Coulomb collisions, background plasma flow, and the
sheath electric field. Two sheath conditions are
considered throughout this work. Under thermal
conditions, the sheath potential is approximated by
the conventional Bohm sheath,

\begin{equation}
V_{\rm sheath}^{\rm th}
\simeq
3\,T_e,
\label{eq:thermal_sheath}
\end{equation}

where $T_e$ is obtained from the SOLPS--ITER plasma
background. Under RF conditions, the sheath
potential is prescribed directly from the
self-consistent COMSOL calculation,

\begin{equation}
V_{\rm sheath}^{\rm RF}
=
V_{\rm RF,DC},
\label{eq:rf_sheath}
\end{equation}

where $V_{\rm RF,DC}$ denotes the rectified RF sheath
potential.

The sheath electric field employed during particle
integration in GITR is evaluated using the Brooks
magnetic pre-sheath formulation
\cite{Brooks:1990}, which provides a first-order
representation of the combined Debye sheath and
magnetic pre-sheath (Chodura sheath). The total
sheath potential is partitioned according to the
local magnetic-field incidence angle,

\begin{equation}
\begin{aligned}
V_{\rm Debye}
&=
f_D(\theta)V_{\rm sheath},
\\
V_{\rm Chodura}
&=
\left[1-f_D(\theta)\right]V_{\rm sheath},
\end{aligned}
\label{eq:brooks_partition}
\end{equation}

where $f_D(\theta)$ is the angle-dependent fraction
of the sheath potential dropping across the Debye
sheath, $\theta$ is the magnetic-field incidence
angle measured relative to the surface normal, and
the remaining potential drop occurs across the
magnetic pre-sheath. The corresponding normal
electric field is

\begin{equation}
\begin{aligned}
E_n(r)
&=
E_{\rm Debye}
+
E_{\rm Chodura},
\\
E_{\rm Debye}
&=
\frac{f_D(\theta)V_{\rm sheath}}
{2\lambda_D}
\exp\!\left(
-\frac{r}{2\lambda_D}
\right),
\\
E_{\rm Chodura}
&=
\frac{\left[1-f_D(\theta)\right]V_{\rm sheath}}
{\rho_i}
\exp\!\left(
-\frac{r}{\rho_i}
\right),
\end{aligned}
\label{eq:brooks_field}
\end{equation}

where $r$ is the wall-normal distance from the
plasma-facing surface, $\lambda_D$ is the local
Debye length, and $\rho_i$ is the local ion
gyroradius. The Debye-sheath contribution dominates
close to the material surface, whereas the magnetic
pre-sheath extends over distances comparable to the
ion gyroradius. Consequently, the spatial extent and
magnitude of the sheath electric field vary
continuously with magnetic-field incidence angle,
producing different ion acceleration histories,
impact energies, and impact angles.

The thermal sheath and particle-acceleration models
implemented in GITR have been independently
benchmarked against the analytical formulation of
Borodkina \textit{et al.}
\cite{Borodkina:2015,Borodkina:2016} for both
electrostatic potential profiles and ion
impact-angle distributions over magnetic-field
incidence angles ranging from normal to nearly
grazing incidence (see \ref{app:gitr_validation}). This benchmark
provides additional confidence in the near-wall
particle trajectories and geometry-specific IEADs
employed throughout the present erosion calculations.

For each ion species and charge state, GITR computes
a unique IEAD on every surface element of the
defeatured ITER ICRH antenna. The spatial variation
of these distributions arises from the combined
influence of the local sheath potential,
magnetic-field incidence angle, plasma density,
temperature, flow velocity, particle charge state,
and antenna geometry. Consequently, neighboring
surface elements may experience substantially
different impact-energy and impact-angle
distributions even when exposed to similar sheath
potentials.

The local IEADs are subsequently convolved with the
corresponding RustBCA sputtering-yield tables,

\begin{equation}
Y_{\rm eff}
=
\int_0^{E_{\max}}
\int_0^{\pi/2}
Y(E,\theta)
f(E,\theta)
\,d\theta\,dE,
\label{eq:Yeff}
\end{equation}

where $Y(E,\theta)$ is the energy- and
angle-dependent sputtering yield obtained from
RustBCA and $f(E,\theta)$ is the normalized IEAD
computed by GITR. The resulting geometry-specific
effective sputtering yield varies from one surface
element to another according to the local IEAD rather
than the sheath potential alone.

\begin{figure*}
\centering
\includegraphics[width=\linewidth]{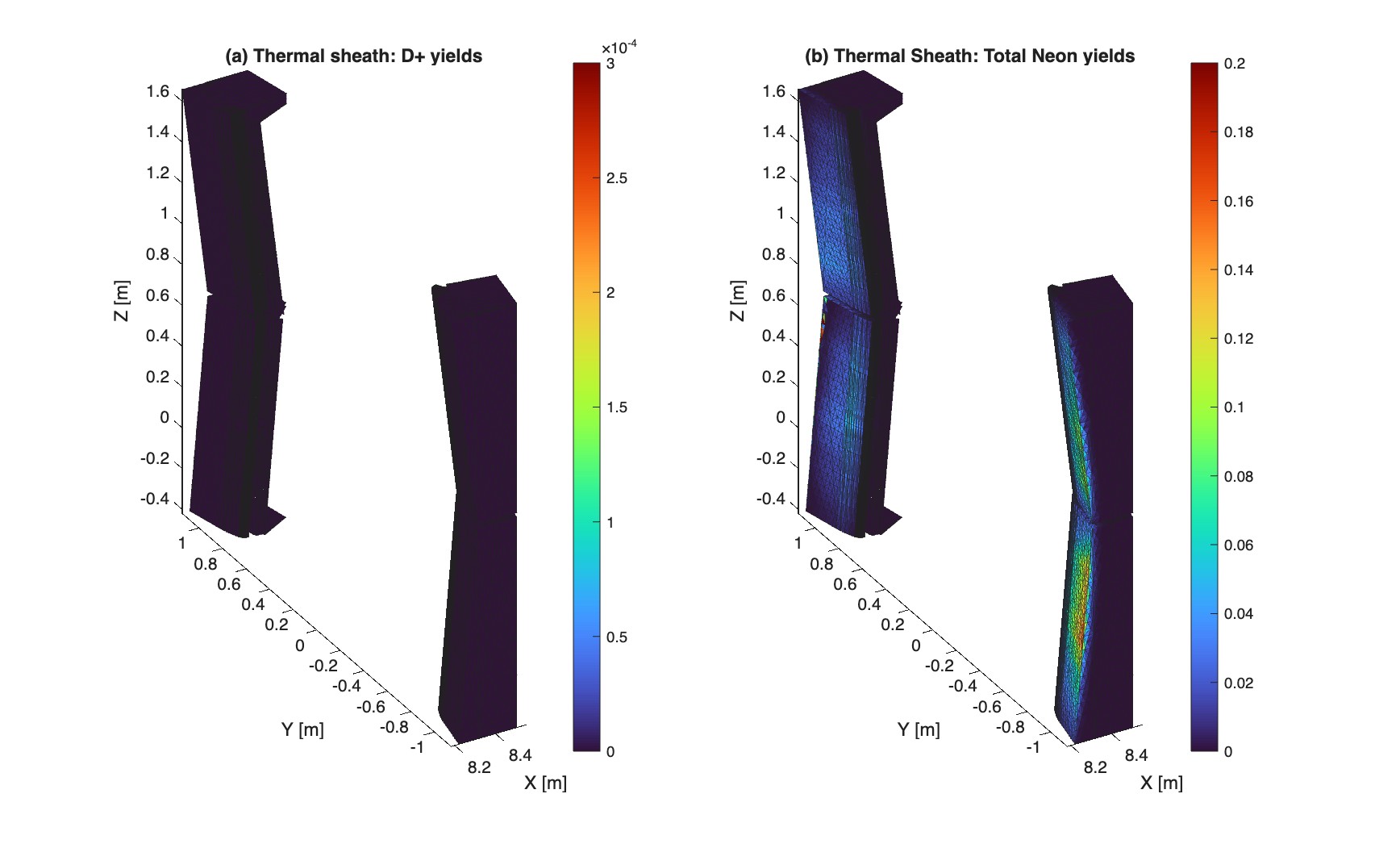}
\caption{
Geometry-specific effective sputtering yields
($Y_{\rm eff}$) under thermal sheath conditions for
(a) D$^{+}$ and (b) the summed contribution from
Ne$^{1+}$--Ne$^{10+}$. The effective sputtering
yields are obtained by convolving the GITR-computed
local IEADs with the RustBCA sputtering-yield
database.
}
\label{fig:iter_effYields_thermal}
\end{figure*}

Figure~\ref{fig:iter_effYields_thermal} presents the
geometry-specific effective sputtering yields under
thermal sheath conditions. The incident ion energy is
controlled primarily by the thermal sheath potential,
such that the characteristic sheath energy scales as
$qV_{\rm sheath}^{\rm th}\simeq3qT_e$. For
D$^{+}$, this energy remains below the W sputtering
threshold of approximately 250--300~eV over nearly
the entire antenna. Consequently, the D$^{+}$
effective sputtering yield is negligible.

Ne possesses a substantially lower W sputtering
threshold of approximately 40~eV and spans charge
states from Ne$^{1+}$ to Ne$^{10+}$. Even under
thermal sheath conditions, the higher charge states
acquire proportionally larger sheath energies,
allowing part of the Ne population to exceed the W
sputtering threshold. The summed Ne effective
sputtering yield reaches approximately
$0.2$ atoms/ion. Its spatial distribution varies
relatively smoothly across the limiter surfaces and
reflects the local electron temperature,
magnetic-field incidence angle, charge-state
composition, and resulting IEADs. The pronounced
spatial localization associated with RF sheath
acceleration is absent.

\begin{figure*}
\centering
\includegraphics[width=\linewidth]{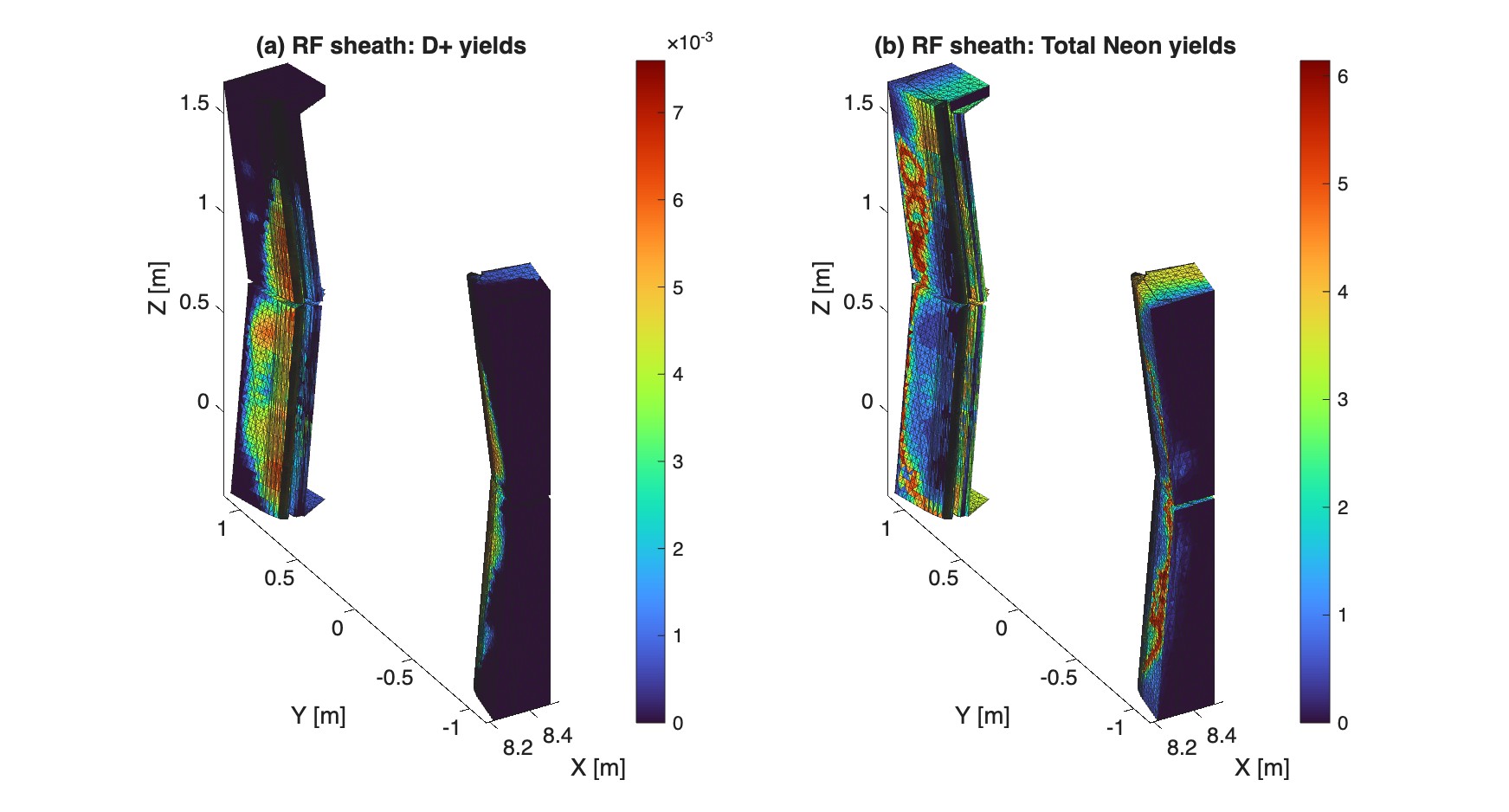}
\caption{
Geometry-specific effective sputtering yields
($Y_{\rm eff}$) under RF sheath conditions for
(a) D$^{+}$ and (b) the summed contribution from
Ne$^{1+}$--Ne$^{10+}$. The effective sputtering
yields are obtained by convolving the GITR-computed
local IEADs with the RustBCA sputtering-yield
database. RF sheath acceleration substantially
modifies the local IEADs, producing strongly
localized effective sputtering yields.
}
\label{fig:iter_effYields_RF}
\end{figure*}

Figure~\ref{fig:iter_effYields_RF} presents the
geometry-specific effective sputtering yields under
RF sheath conditions. The rectified RF sheath
substantially increases the ion impact energies and
modifies the impact-angle distributions relative to
the thermal sheath case. Because the resulting IEADs
depend on the local sheath potential,
magnetic-field incidence angle, particle charge state,
plasma conditions, and antenna geometry, the RF
effective-yield distributions exhibit pronounced
spatial localization.

For D$^{+}$, the effective sputtering yield increases
by more than an order of magnitude, reaching
approximately $8\times10^{-3}$ atoms/ion. Because D
is singly charged, the energy gained across the RF
sheath is limited to approximately
$V_{\rm RF,DC}$. The D$^{+}$ IEADs are therefore
primarily controlled by the local RF sheath
acceleration, and the spatial distribution of
$Y_{\rm eff}$ broadly follows the RF sheath
potential. Regions of enhanced D$^{+}$ effective
yield consequently coincide with regions of elevated
sheath potential.

The Ne response is more complex because the impurity
population spans charge states from Ne$^{1+}$ to
Ne$^{10+}$. Each charge state acquires a different
energy across the RF sheath and produces a distinct
local IEAD. The charge states also exhibit different
gyroradii, collisional responses, and finite-orbit
effects, resulting in charge-state-dependent impact
locations and incidence angles. Summing these
contributions produces a broader effective-yield
distribution than that obtained for D$^{+}$.

The maximum summed Ne effective sputtering yield
exceeds $Y_{\rm eff}\approx6$ atoms/ion and occurs
near the outer portions of the limiter faces rather
than directly adjacent to the Faraday-screen bars,
where the RF sheath potential is largest. This
spatial offset demonstrates that the effective
sputtering yield is not determined by the sheath
potential alone. Instead, it reflects the complete
charge-state-dependent IEAD, including the impact
energy, incidence angle, and particle trajectory
through the local electric and magnetic fields.

Comparison of the D and Ne distributions therefore
shows that the spatial pattern of $Y_{\rm eff}$ is
species dependent. The D$^{+}$ yield map largely
follows the sheath-voltage structure because its IEAD
is governed by singly charged RF acceleration. In
contrast, the summed Ne distribution reflects the
superposition of multiple charge-state-resolved IEADs
with different impact energies, incidence angles, and
surface-impact locations.

Overall, Ne produces effective sputtering yields
nearly three orders of magnitude larger than D$^{+}$
over much of the antenna surface. The resulting
geometry-specific $Y_{\rm eff}$ distributions are
subsequently combined with the corresponding
charge-state-resolved incident ion fluxes to calculate
the gross tungsten erosion source presented in the
following section.

\subsection{Local Tungsten Erosion Under Thermal and RF Sheath Conditions}
\label{sec:5.4}

Using the geometry-specific effective sputtering yields,
$Y_{\rm eff}(\mathbf{s})$,
together with the local ion fluxes obtained from extrapolated $n_s \times v_s$ (here \emph{s} stands for species) and the GITR-computed IEADs, the gross W erosion flux at each surface element $\mathbf{s}$ is computed as

\begin{equation}
\Gamma_{\rm gross,W}(\mathbf{s})
=
\sum_{i,q}
Y_{\rm eff}^{(i,q)}(\mathbf{s})
\Gamma_{\rm ion}^{(i,q)}(\mathbf{s}),
\label{eq:gross_erosion}
\end{equation}

where $i$ denotes the projectile species, $q$ the corresponding charge state, and
$\Gamma_{\rm ion}^{(i,q)}(\mathbf{s})$
is the local ion flux incident on the surface.

To quantify the influence of RF sheath acceleration, W erosion is evaluated under both thermal and RF sheath conditions using the same SOLPS-ITER plasma background. Under thermal conditions, the incident ion energy is determined by the conventional Bohm sheath
($V_{\rm sheath}\approx3T_e$),
whereas under RF conditions it is determined by the COMSOL-computed rectified RF sheath potential together with the charge-state-dependent RF acceleration.

\begin{figure}
\centering
\includegraphics[width=\linewidth]{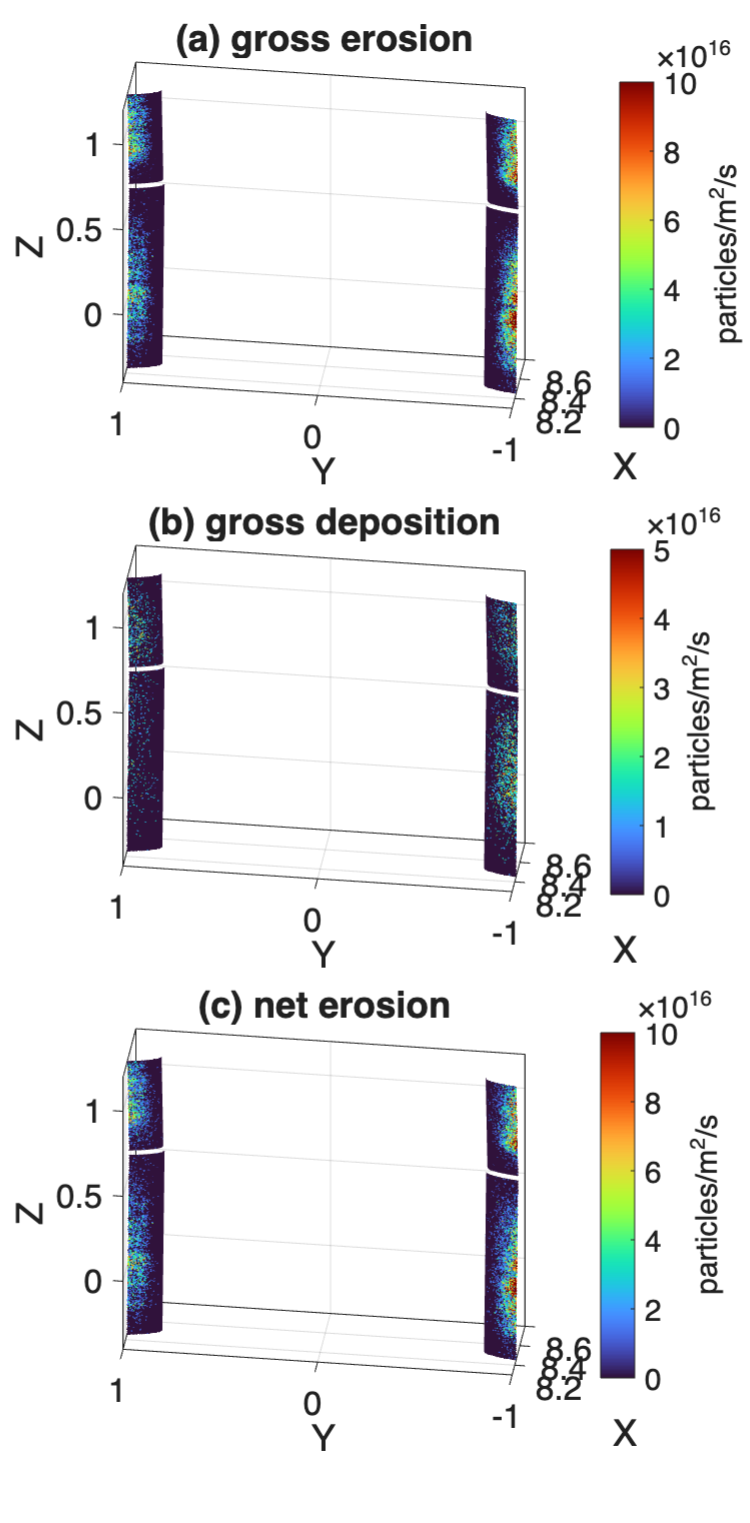}
\caption{
Three-dimensional W erosion predicted under thermal sheath conditions showing (a) gross erosion, (b) gross deposition, and (c) net erosion. Units are particles~m$^{-2}$~s$^{-1}$.
}
\label{fig:iter_ero_thermal}
\end{figure}

Figure~\ref{fig:iter_ero_thermal} presents the three-dimensional distributions of gross erosion, gross deposition, and net erosion predicted under thermal sheath conditions. The corresponding area-integrated gross erosion, gross deposition, and net erosion rates are
$5.21\times10^{16}$,
$5.33\times10^{15}$,
and
$4.68\times10^{16}$~particles~s$^{-1}$,
respectively. Consequently, only approximately 10\% of the thermally sputtered W is redeposited onto nearby limiter surfaces, while nearly 90\% escapes the immediate erosion region.

Under thermal sheath conditions, the ion impact energy is determined primarily by the Bohm sheath,
$qV_{\rm sheath}^{\rm th}\approx3qT_e$,
such that the sputtering yield follows the relatively smooth spatial variation of the background edge plasma. As shown in Fig.~\ref{fig:iter_ero_thermal}(a), the highest gross erosion occurs along the plasma-facing leading edges of the antenna limiters where both the electron temperature and incident plasma flux are largest. Moving away from these exposed regions, the decrease in plasma temperature and particle flux rapidly lowers the incident ion energy below the efficient sputtering regime, resulting in only weak erosion over the recessed limiter surfaces. Consequently, the thermal erosion footprint remains confined primarily to the leading plasma-facing edges and exhibits relatively smooth spatial variation without strong localized structure.

The corresponding gross deposition and net erosion distributions shown in Figs.~\ref{fig:iter_ero_thermal}(b,c) closely resemble the gross erosion pattern. Because sputtering originates almost entirely from the leading limiter edges, redeposition likewise remains localized in their vicinity and only partially compensates the local material loss. As a result, the net erosion distribution differs only modestly from the gross erosion profile, reflecting the relatively small local redeposition fraction under thermal sheath conditions.

\begin{figure}
\centering
\includegraphics[width=\linewidth]{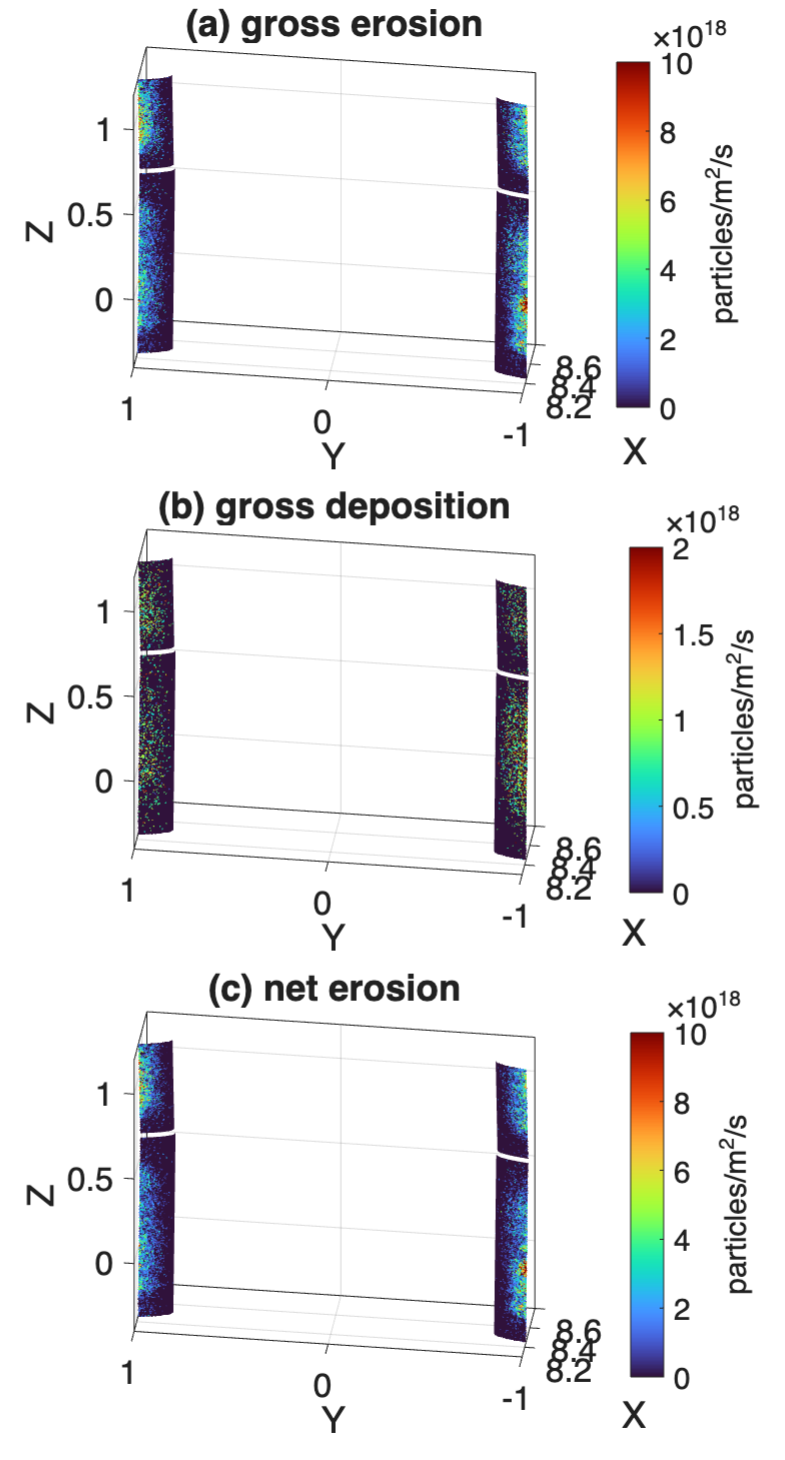}
\caption{
Three-dimensional W erosion predicted under RF sheath conditions showing (a) gross erosion, (b) gross deposition, and (c) net erosion. Compared with the thermal sheath case shown in Fig.~\ref{fig:iter_ero_thermal}, RF sheath acceleration increases the W source and broadens the erosion footprint over the limiter surfaces. Units are particles~m$^{-2}$~s$^{-1}$.
}
\label{fig:iter_ero_3d}
\end{figure}

The situation changes substantially under RF operation. Figure~\ref{fig:iter_ero_3d} presents the corresponding gross erosion, gross deposition, and net erosion distributions obtained using the COMSOL-computed RF sheath potentials. RF sheath acceleration increases the area-integrated gross erosion to
$3.34\times10^{18}$~particles~s$^{-1}$,
corresponding to an enhancement of approximately a factor of 64 relative to the thermal sheath case. The corresponding gross deposition and net erosion rates become
$3.33\times10^{17}$
and
$3.01\times10^{18}$~particles~s$^{-1}$,
respectively, while the redeposition fraction remains close to 10\%.

Unlike the thermal sheath case, the incident ion energy is no longer determined solely by the local electron temperature. Instead, ions experience additional acceleration through the spatially varying RF sheath potential, with the resulting sputtering governed by the charge-state-dependent RF-modified IEADs. 

Consequently, the erosion morphology no longer follows only the background plasma profiles. Comparison of Figs.~\ref{fig:iter_ero_thermal}(a) and~\ref{fig:iter_ero_3d}(a) shows that RF sheath acceleration extends the erosion to an extent poloidally and across the central portions of the limiter where RF acceleration raises the ion energies above the sputtering threshold. The resulting erosion footprint is therefore a bit broader and more spatially distributed than in the thermal sheath case.


An important observation is that the locations of
maximum erosion do not coincide directly with the
regions of highest RF sheath voltage adjacent to the
FS bars. Instead, the strongest erosion occurs where
elevated RF sheath potentials overlap with
sufficiently large incident plasma fluxes along the
exposed limiter surfaces. Regions experiencing large
RF sheath potentials (\emph{i.e.}, close to the FS
bars) but comparatively weak plasma flux exhibit
negligible erosion, whereas locations combining
energetic RF-accelerated ions with high particle flux
produce the largest sputtering rates. These results
demonstrate that the local tungsten source is
determined by the combined influence of RF sheath
acceleration and incident plasma flux rather than by
either quantity independently. Consequently, the
regions of largest RF sheath potential associated
with the vicinity of the $S=0$ resonance-cone
structure do not provide the dominant contribution to
the gross W erosion under the present ITER
conditions. This observation reduces the sensitivity
of the predicted erosion source to the uncertainty in
the absolute RF sheath amplitudes near the $S=0$
region, where cold-plasma RF sheath models are known
to overestimate the rectified sheath
potential.

The gross deposition distributions shown in Figs.~\ref{fig:iter_ero_thermal}(b) and~\ref{fig:iter_ero_3d}(b) exhibit a similar evolution. Under thermal sheath conditions, deposition remains weak and localized near the leading limiter edges where sputtering originates. Under RF sheath conditions, the deposition footprint expands together with somewhat broader erosion source; however, its magnitude remains substantially smaller than the corresponding gross erosion over the entire antenna structure. Consequently, the net erosion distributions shown in Figs.~\ref{fig:iter_ero_thermal}(c) and~\ref{fig:iter_ero_3d}(c) closely follow the corresponding gross erosion patterns, indicating that local redeposition removes only a relatively small fraction of the sputtered material.

The relatively small redeposition fraction reflects the plasma conditions surrounding the ITER ICRH antenna. Under the present low-density, low-$T_e$ scrape-off-layer conditions, sputtered W neutrals possess ionization mean free paths of several metres, allowing many particles to leave the immediate erosion region before becoming ionized. Following ionization, the resulting W ions become strongly magnetized, with Larmor radii of only a few millimetres, and are transported predominantly along magnetic field lines. The combination of long neutral free flight and subsequent magnetic confinement therefore produces finite but incomplete local redeposition under both thermal and RF sheath conditions.

\subsection{Area-Integrated Erosion Balance}
\label{sec:5.5}

To quantify the influence of RF sheath acceleration
on W sputtering, the erosion source is integrated over
the total sheath-active limiter area,
$A_{\rm lim}=1.383~{\rm m^2}$.
Table~\ref{tab:erosion_balance} summarizes the
resulting erosion balance under thermal and RF sheath
conditions.

\begin{table}
\caption{
Area-integrated W erosion balance over the
sheath-active limiter area under thermal and RF
sheath conditions.
}
\centering
\begin{tabular}{lcc}
\hline
Quantity & Thermal & RF \\
\hline
Gross erosion ($\mathrm{s^{-1}}$)
& $5.21\times10^{16}$
& $3.34\times10^{18}$ \\

Gross deposition ($\mathrm{s^{-1}}$)
& $5.33\times10^{15}$
& $3.33\times10^{17}$ \\

Net erosion ($\mathrm{s^{-1}}$)
& $4.68\times10^{16}$
& $3.01\times10^{18}$ \\

Redeposition fraction
& $\sim10\%$
& $\sim10\%$ \\

W self-sputtering ($\mathrm{s^{-1}}$)
& $9.88\times10^{14}$
& $1.82\times10^{17}$ \\

Self-sputtering fraction
& 1.9\%
& 5.5\% \\

Gross erosion enhancement
& 1
& $\times64$ \\
\hline
\end{tabular}
\label{tab:erosion_balance}
\end{table}

The integrated erosion balance demonstrates that RF
sheath acceleration increases the gross W erosion
source by approximately a factor of 64 while producing
only a minor change in the local redeposition
fraction. Under thermal sheath conditions,
approximately 10.2\% of the sputtered W is
redeposited on nearby plasma-facing components,
compared with approximately 10.0\% under RF sheath
conditions. Thus, the principal effect of RF sheath
acceleration is to increase the magnitude of the W
source rather than to substantially modify the
fraction of sputtered material that is locally
redeposited.

\begin{figure}
\centering
\includegraphics[width=0.95\linewidth]
{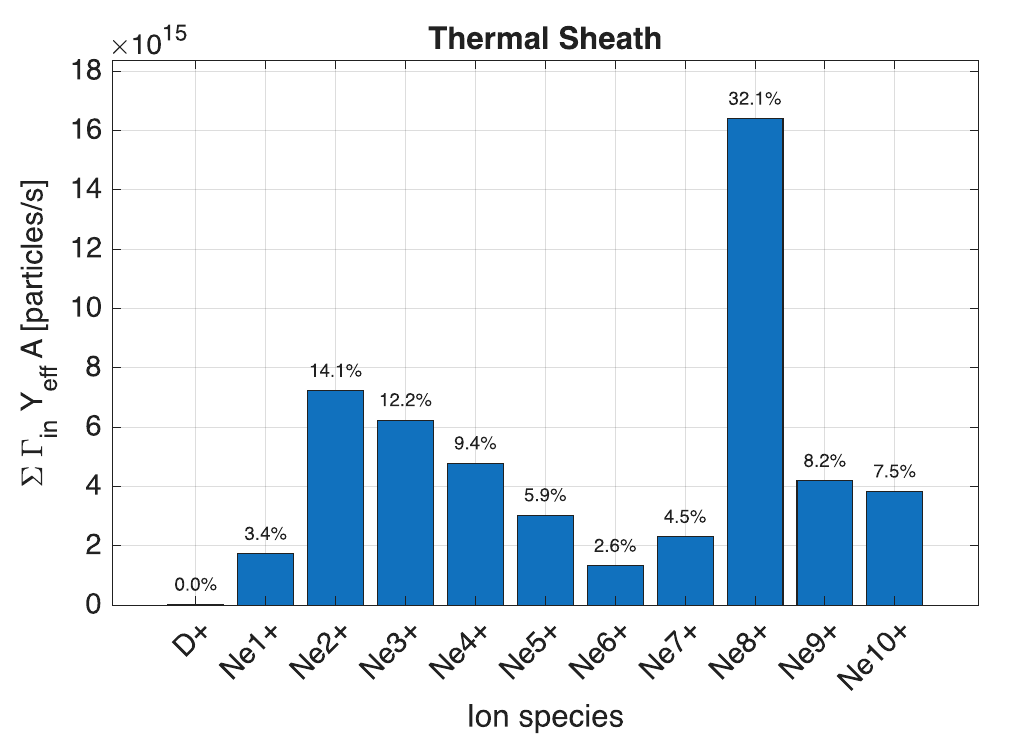}
\caption{
Area-integrated primary plasma-ion W sputtering
source under thermal sheath conditions, resolved by
incident species and charge state. Percentages denote
the contributions from D and
Ne$^{1+}$--Ne$^{10+}$ and are normalized to 100\%
after excluding W self-sputtering. The corresponding
W self-sputtering contribution is reported separately
in Table~\ref{tab:erosion_balance}.
}
\label{fig:int_flux_thermal}
\end{figure}

\begin{figure}
\centering
\includegraphics[width=0.95\linewidth]
{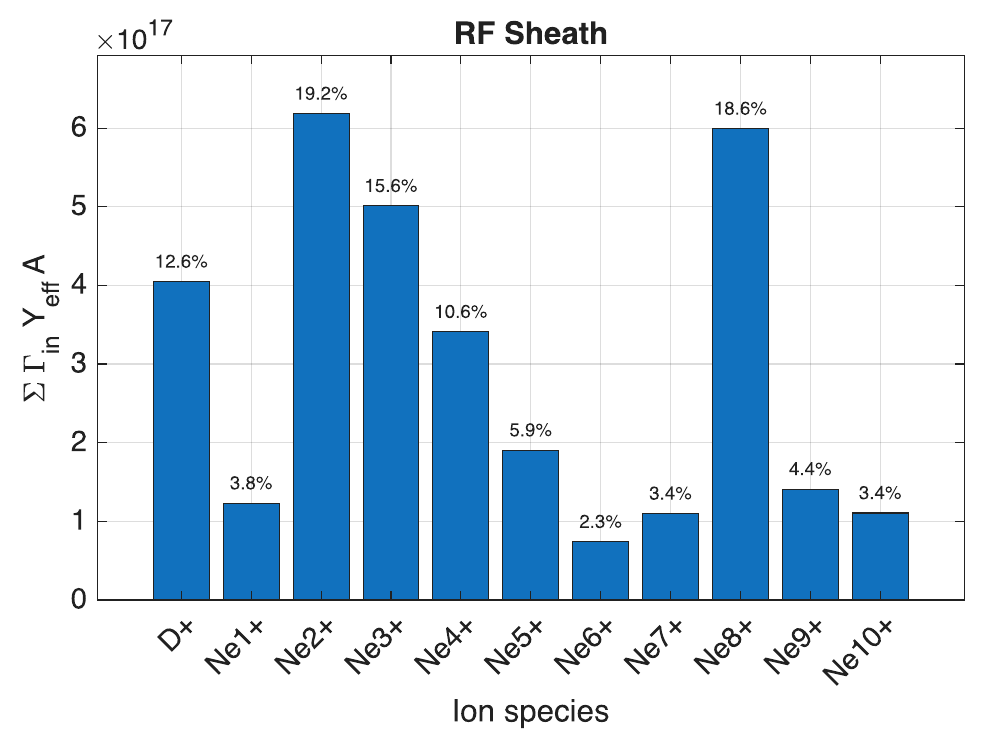}
\caption{
Area-integrated primary plasma-ion W sputtering
source under RF sheath conditions, resolved by
incident species and charge state. Percentages denote
the contributions from D and
Ne$^{1+}$--Ne$^{10+}$ and are normalized to 100\%
after excluding W self-sputtering. The corresponding
W self-sputtering contribution is reported separately
in Table~\ref{tab:erosion_balance}.
}
\label{fig:int_flux_RF}
\end{figure}

The area-integrated W self-sputtering source also
increases substantially under RF operation, rising
from $9.88\times10^{14}$~particles~s$^{-1}$ under
thermal sheath conditions to
$1.82\times10^{17}$~particles~s$^{-1}$ under RF
sheath conditions. These values correspond to 1.9\%
and 5.5\%, respectively, of the total gross W erosion
rate originating from the ICRH antenna limiters. Figures~\ref{fig:int_flux_thermal} and
\ref{fig:int_flux_RF}, however, show only the
charge-state-resolved contributions to the
\emph{primary plasma-ion sputtering source},
comprising incident D and Ne ions. The percentages
shown in these figures are therefore normalized to
100\% after excluding W self-sputtering, which is
reported separately in Table~\ref{tab:erosion_balance}.
The primary plasma-ion source remains dominated by
Ne under both sheath conditions, while W
self-sputtering provides a smaller secondary
contribution and remains well below the level required
to produce a self-sustaining sputtering cascade. This behavior results primarily from the relatively
low prompt redeposition fraction and the long
ionization mean free paths under the present ITER
edge conditions, which transport most sputtered W
away from the RF sheath-active limiter surfaces
before it can participate in repeated self-sputtering.

Figure~\ref{fig:int_flux_thermal} presents the
area-integrated primary plasma-ion W sputtering source
under thermal sheath conditions, resolved by
projectile species and charge state. Because the
available thermal sheath energy,
$qV_{\rm sheath}^{\rm th}\approx3qT_e$, remains below
the sputtering threshold for D on W, the deuterium
contribution is negligible, accounting for only 0.04\%
of the primary plasma-ion source. Sputtering is
instead almost entirely impurity driven, with Ne
contributing more than 99\% of the primary
plasma-ion W source.

Among the individual Ne charge states,
Ne$^{8+}$ provides the largest contribution
(32.1\%), followed by Ne$^{2+}$ (14.1\%),
Ne$^{3+}$ (12.2\%), and Ne$^{4+}$ (9.4\%).
These results demonstrate that, even in the absence
of RF sheath acceleration, the comparatively low
sputtering threshold of Ne on W, together with the
local Ne charge-state distribution and incident
particle flux, is sufficient for Ne to dominate the
primary W sputtering source.

Figure~\ref{fig:int_flux_RF} presents the
corresponding primary W sputtering source
under RF sheath conditions based on the COMSOL calculations presented in this work. RF sheath acceleration
raises the D ion energies above the W sputtering
threshold, increasing the D contribution from
essentially zero under thermal conditions to
approximately 12.6\% of the primary W
source. Nevertheless, Ne remains dominant,
collectively contributing approximately 87.4\% of
the primary plasma-ion source.

Among the individual Ne charge states,
Ne$^{2+}$ (19.2\%) and Ne$^{8+}$ (18.6\%) provide
the largest contributions, followed by Ne$^{3+}$
(15.6\%) and Ne$^{4+}$ (10.6\%). Together, these
four charge states account for more than 60\% of the
primary plasma-ion W sputtering source. A
qualitatively similar behavior was observed in our
previous RF-induced erosion studies of oxygen
impurities in WEST, where intermediate charge states
also provided the dominant contribution to the
antenna sputtering source despite the presence of
more highly ionized species~\cite{Klepper:2022,kumar:2025,kumar_rfppc_2025}.

An important observation is that the highest Ne
charge states do not necessarily dominate the
integrated erosion source. Although Ne$^{9+}$ and
Ne$^{10+}$ acquire the largest energies through RF
sheath acceleration, their comparatively smaller
incident fluxes limit their overall contributions.
Conversely, intermediate charge states such as
Ne$^{2+}$, Ne$^{3+}$, Ne$^{4+}$, and Ne$^{8+}$
combine sufficiently strong RF sheath acceleration
with substantially larger incident particle fluxes,
making them the dominant contributors. These results
further demonstrate that RF-induced W erosion is
governed by the combined effects of RF-modified IEADs
and incident plasma flux rather than by RF sheath
acceleration alone.

For the same wide-grid SOLPS-ITER plasma
background (IMAS case~\#123361), recent ERO2.0
simulations predict gross thermal W erosion rates of
approximately
$6.8\times10^{21}$~atoms~s$^{-1}$ from the
divertor and
$9.9\times10^{20}$~atoms~s$^{-1}$ from the
main chamber during Ne-seeded
operation~\cite{Baumann2026}. The RF-specific gross
W source identified here
($3.34\times10^{18}$~atoms~s$^{-1}$) is therefore
approximately three orders of magnitude smaller than
the divertor source and at least two orders of magnitude
smaller than the thermal main-chamber source.
Nevertheless, unlike the dominant divertor source,
the RF-induced W source originates from localized
low-field-side main-chamber ICRH structures and is
therefore expected to experience transport pathways
more similar to those of the thermal main-chamber W
source than to divertor-generated tungsten. Recent
ERO2.0 simulations further indicate that the thermal
main-chamber W source originates predominantly from
the low-field side and, together with the baffle top,
constitutes an important contributor to the tungsten
influx toward the confined
plasma~\cite{Baumann2026}. Consequently, despite its
comparatively small magnitude, the localized
RF-generated antenna source may exhibit transport
characteristics distinct from those of
divertor-generated tungsten, providing additional
motivation for the three-dimensional transport and W core-contamination assessment presented in Section~\ref{sec:6}.

\section{Global Transport of Tungsten Impurities}
\label{sec:6}

\begin{figure}
\centering
\includegraphics[width=\linewidth]{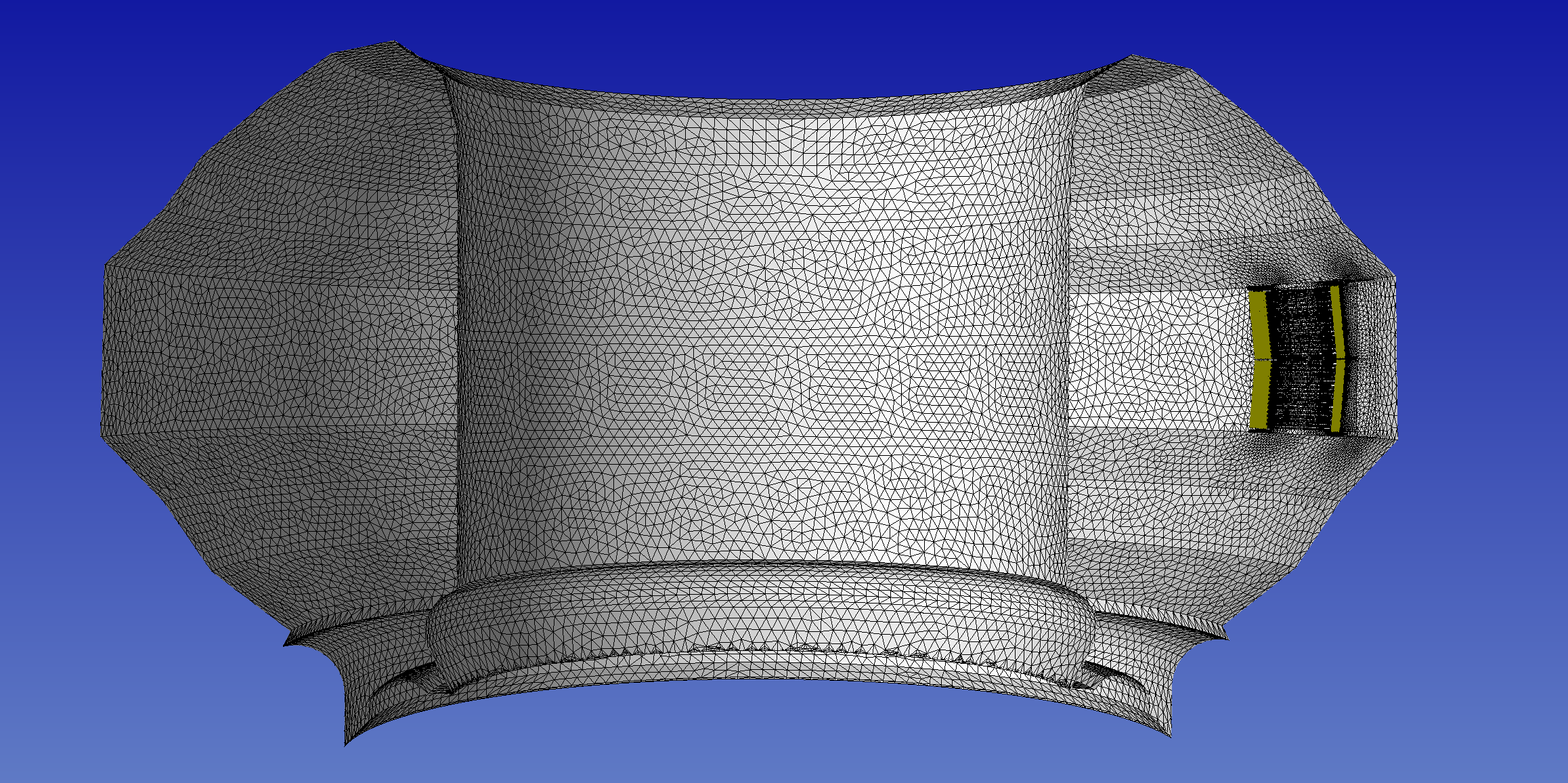}
\caption{
Highly graded three-dimensional unstructured GITRm
mesh used for the ITER global W transport
calculations. Local mesh refinement is applied around
the ICRH antenna and limiter structures to resolve
localized erosion, redeposition, and near-field
impurity transport.
}
\label{fig:gitrm_mesh}
\end{figure}

The global transport of W originating from the ITER
ICRH antenna is simulated using GITRm on a fully
three-dimensional unstructured mesh representing the
ITER first wall, ICRH antenna, limiter structures,
and divertor. Additional details of the transport model
and numerical implementation are provided in \ref{app:gitrm}. Figure~\ref{fig:gitrm_mesh}
shows the computational mesh employed in the present
simulations. Local refinement is applied around the
antenna aperture and limiter structures to resolve the
localized RF-driven W source, whereas a progressively
coarser mesh is used elsewhere to efficiently capture
impurity transport throughout the ITER vessel.

Unlike the local erosion analysis, the global transport calculation is performed only for the RF sheath case. The impurity evolution is treated in the trace-impurity approximation, in which the prescribed background plasma, electromagnetic fields, and transport coefficients remain independent of the evolving W density, while W--W collisions are neglected. The initial marker distribution is sampled from the spatially resolved RF-driven gross erosion source predicted by the STRIPE workflow. The marker weights are normalized to the corresponding physical W source rate, allowing the GITRm density tally to be converted into a physical W inventory. Following initialization, sputtered W neutrals undergo ionization and recombination using ADAS rate coefficients, together with Lorentz acceleration, Coulomb collisions, anomalous cross-field diffusion, and repeated interactions with PFCs, including W self-sputtering. At the antenna surface, these subsequent W--surface interactions account for acceleration by the rectified RF sheath potential. 

Following sputtering, W transport proceeds through
two distinct stages. Initially, sputtered atoms are
emitted as energetic neutrals and propagate
ballistically without being influenced by the magnetic
field. Following ionization, the resulting W ions
become strongly magnetized and are transported
predominantly along magnetic field lines. The
competition between these two regimes determines the
balance among local redeposition, scrape-off-layer
transport, and penetration into the confined plasma.

\begin{figure}
\centering
\includegraphics[width=\linewidth]
{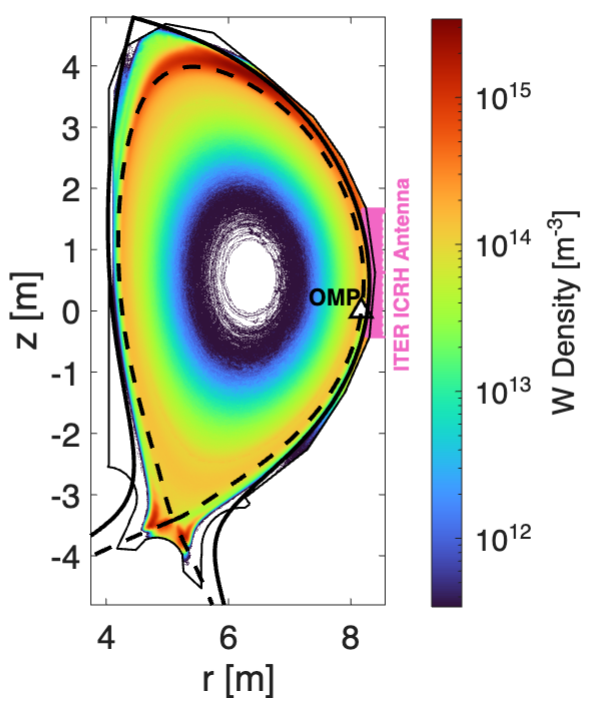}
\caption{
Tungsten impurity transport predicted by GITRm after
100~ms of evolution following a single initialization
of the RF-generated sputtered W source. The
three-dimensional density distribution and
corresponding poloidal projection demonstrate the
formation of an antenna-connected impurity channel
extending through the scrape-off layer toward the
divertor. The W density is tallied on the GITRm grid,
whereas the concentration reported in
Eq.~\eqref{eq:cWcore} is evaluated only over the
SOLPS-covered annular confined-plasma region,
$0.88\leq\psi_N\leq1.00$, where a finite mapped
electron density is available. Cells with zero tallied
W density are omitted from the logarithmic color
scale and therefore appear blank in the poloidal
projection.
}
\label{fig:iter_transport}
\end{figure}

Figure~\ref{fig:iter_transport} presents the W density
distribution after 100~ms of evolution following a
single initialization of the RF-generated sputtered W
source. The calculation predicts the formation of an
antenna-connected impurity reservoir extending along
open magnetic field lines. Following ionization, the
sputtered W is transported predominantly through the
SOL, producing a continuous impurity channel
extending from the ICRH antenna toward the divertor.
Although the highest W densities remain localized
near the antenna, finite impurity concentrations extend
several metres from the source region and approximately
follow the magnetic flux surfaces, demonstrating
efficient field-aligned transport throughout the SOL.

Under the ITER edge conditions considered here, the
relatively low plasma density and electron temperature
surrounding the ICRH antenna produce W ionization
mean free paths of several metres. Consequently, many
sputtered neutrals escape the immediate erosion region
before becoming ionized, consistent with the
comparatively small prompt redeposition fraction
reported in Table~\ref{tab:erosion_balance}. Once
ionized, however, the corresponding Larmor radii
decrease to only a few millimetres, causing the
impurities to become tightly magnetized and
transported predominantly along open magnetic field
lines. The transition from ballistic neutral transport
to field-aligned ion transport therefore explains both
the antenna-connected impurity reservoir and the
continuous SOL transport pathway linking the antenna
to the divertor despite the localized erosion source.

To quantify the global transport, the physical W
inventory represented by the source-normalized GITRm
density tally is integrated over the computational
domain after 100~ms of evolution. At this time, the
total mobile W inventory remaining on the GITRm
grid is $1.30\times10^{17}$ particles. Of this
inventory, $2.86\times10^{16}$ particles are located
inside the separatrix within the confined-plasma
region covered by the wide-grid SOLPS-ITER mesh.
Thus, approximately 22.0\% of the mobile W
inventory remaining in the plasma at 100~ms lies
within the SOLPS-covered confined-plasma region.

This fraction is conditional on the W population
remaining in the GITRm plasma-density tally at
$t=100$~ms. W particles that have intersected
material surfaces and been recorded as deposited are
not included in either the numerator or denominator.
The 22.0\% value should therefore not be interpreted
as the fraction of the complete initially launched W
population that enters the confined plasma. Rather, it
describes the spatial partitioning of the mobile W
inventory remaining in the plasma after 100~ms. The
deposited population is accounted for separately
through the surface-interaction tallies used to
calculate the gross deposition and net erosion rates.

The wide-grid SOLPS-ITER solution does not extend to
the magnetic axis. The innermost location inside the
separatrix containing a finite mapped electron
density occurs at approximately
$\psi_N=0.86$. To minimize interpolation and boundary
effects, the W concentration is evaluated only over
the SOLPS-covered annular confined-plasma region
$0.88\leq\psi_N\leq1.00$, where finite electron
density is available. The corresponding annular W
concentration is defined as

\begin{equation}
c_{\mathrm{W,ann}}
=
\frac{\displaystyle\int n_{\mathrm W}\,dV}
     {\displaystyle\int n_e\,dV}
=
1.70\times10^{-6},
\label{eq:cWcore}
\end{equation}

where both integrals are evaluated over the same
SOLPS-covered annular confined-plasma region.

Within this region, the integrated W and electron
inventories are $2.86\times10^{16}$ and
$1.68\times10^{22}$ particles, respectively,
corresponding to volume-averaged densities of
$\langle n_{\mathrm W}\rangle
=1.37\times10^{14}$~m$^{-3}$ and
$\langle n_e\rangle
=8.07\times10^{19}$~m$^{-3}$. Their ratio is
consistent with the inventory-based value given by
Eq.~\eqref{eq:cWcore}. The predicted W density
remains several orders of magnitude below the
background plasma density throughout the simulation
domain, supporting the validity of the trace-impurity
approximation adopted in the present GITRm
calculations.

To verify that the reported concentration is
independent of the computational mesh, the
integration was repeated directly on the SOLPS-ITER
grid after interpolating the GITRm W density onto
the SOLPS mesh. This independent evaluation gives
$c_{\mathrm{W,ann}}=1.697\times10^{-6}$ together
with the same 22.0\% fraction of the mobile W
inventory remaining in the plasma within the
SOLPS-covered confined-plasma region, demonstrating
that the reported annular concentration is
insensitive to the choice of integration mesh.

Because the mapped SOLPS-ITER plasma background does
not extend to the magnetic axis, the present
calculation does not provide a volume-averaged W
concentration over the complete ITER confined plasma.
The reported value therefore represents only the
SOLPS-covered annular confined-plasma region
($0.88\leq\psi_N\leq1.00$). A quantitative
whole-core concentration would require extending the
background plasma together with additional core
transport physics, including neoclassical
convection, impurity pinch, and turbulent transport,
and is therefore beyond the scope of the present
work.

The reported inventory likewise corresponds to a
finite-time transport calculation following a single
initialization of the RF-generated W source. The
100~ms transport interval was selected to
characterize the finite-time migration of
antenna-generated W rather than to obtain a
steady-state impurity inventory. While the local
gross erosion, gross deposition, and net erosion
rates stabilize rapidly following the initial
transient, the global W inventory continues to evolve
through ongoing transport and redistribution
throughout the ITER edge plasma. A quantitative
steady-state prediction would require continuous
source injection together with substantially longer
transport simulations. Under the present numerical
configuration, a single 100~ms GITRm calculation
requires approximately one month of wall-clock time
using four NVIDIA A100 (80 GB) GPUs.

These results demonstrate that a finite fraction of
the mobile antenna-generated W inventory reaches the
SOLPS-covered confined-plasma region within 100~ms.
However, the present calculations do not determine
either the whole-core W concentration or its
long-time steady-state accumulation.

\section{Discussion}
\label{sec:7}

The present study provides a scenario-specific predictive assessment of RF-driven plasma--material interaction at the ITER ICRH antenna by coupling the edge-plasma background, RF wave propagation, RF sheath formation, sputtering physics, and three-dimensional impurity transport within the STRIPE framework. Although these processes have previously been investigated individually, the present work demonstrates that the RF-induced W source and its subsequent migration emerge from their combined treatment. In particular, the calculations show that RF sheath formation, impurity charge-state evolution, local plasma conditions, surface sputtering, and subsequent transport collectively determine both the magnitude and spatial distribution of antenna-generated W in the plasma.

A central result of this work is that the RF sheath voltage alone is not a sufficient predictor of W erosion. The rectified RF sheath determines ion acceleration and consequently modifies the local IEADs, whereas the incident plasma flux determines the number of energetic ions reaching the surface. The largest erosion therefore develops where energetic RF-modified IEADs overlap with sufficiently large incident plasma fluxes, rather than where the RF sheath voltage itself is largest. Under the plasma conditions and COMSOL configuration considered here, the maximum RF sheath voltages occur primarily along the recessed limiter sidewalls where the slow wave remains propagative, whereas the largest incident plasma fluxes occur farther along the exposed limiter faces. Consequently, RF-driven PMI is governed by the coupled effects of RF sheath acceleration and local edge-plasma conditions rather than by the RF sheath potential alone.

The comparison between thermal- and RF-sheath conditions further illustrates the transition from conventional PMI to RF-specific sputtering. Under thermal-sheath conditions, the ion impact energies remain limited by the Bohm sheath, and D-induced W sputtering is negligible over most of the antenna because the corresponding energies remain below the W sputtering threshold. RF sheath acceleration raises the impact energies of all ion species and charge states and produces charge-state-dependent impact distributions. The relative enhancement in W sputtering can be particularly important for lower Ne charge states, while highly ionized Ne may already produce substantial sputtering yields under thermal-sheath conditions. As a result, RF operation increases the gross W erosion source by approximately a factor of 64 relative to the thermal-sheath case.

The dominant sputtering mechanism nevertheless remains impurity driven. Under RF-sheath conditions, D contributes approximately 12.6\% of the primary plasma-ion sputtering source, whereas Ne contributes approximately 87.4\%. The dominant Ne contributions arise from Ne$^{2+}$, Ne$^{8+}$, Ne$^{3+}$, and Ne$^{4+}$ rather than exclusively from the highest charge states. This behavior reflects the balance between charge-dependent sheath acceleration and incident plasma flux. Although Ne$^{9+}$ and Ne$^{10+}$ experience the largest energy gain across the sheath, their smaller incident fluxes limit their integrated contribution. By contrast, intermediate charge states combine sufficiently strong acceleration with larger local particle fluxes. W self-sputtering provides an additional secondary contribution but accounts for only approximately 5.5\% of the total gross erosion under RF-sheath conditions and does not produce a self-sustaining sputtering cascade.

The present calculations also demonstrate the importance of resolving realistic IEADs when evaluating RF-driven erosion. Conventional estimates frequently relate sputtering directly to the local RF sheath potential or assume monoenergetic ion bombardment. Neither approximation is adequate for the realistic ITER antenna limiter geometry considered here. Each ion species and charge state experiences a distinct acceleration history determined by the local sheath potential, magnetic-field incidence angle, Coulomb collisions, plasma flow, gyromotion, finite-orbit effects, and antenna geometry. The effective sputtering yield therefore results from the convolution of the local IEAD with the corresponding energy- and angle-dependent RustBCA yield. Consequently, even neighboring limiter surface elements exposed to similar RF sheath voltages may exhibit substantially different erosion because of differences in their impact-energy distributions, impact angles, and incident plasma fluxes. The present analysis is restricted to the RF-sheath-active limiter surfaces; whether similar behavior occurs on neighboring first-wall panels or other plasma-facing components experiencing RF-rectified fields remains beyond the scope of the present study.

The quantitative erosion predictions inherit uncertainties associated with both the sputtering and RF-sheath models. The density extrapolation procedure used inside the recessed antenna region likely overestimates the local plasma density and therefore provides a conservative estimate of the RF-specific W source under the assumptions considered here. RustBCA employs the binary-collision approximation and does not include many-body interactions, dynamic surface evolution, surface roughness, or mixed-material effects. Its accuracy is therefore expected to decrease near the sputtering threshold and for surface conditions that differ substantially from an ideal W target. Furthermore, the experimental database for W sputtering is concentrated primarily at normal incidence, whereas the present calculations rely on angle-dependent RustBCA predictions over a broad range of impact angles.

The RF-sheath calculations employ the sheath-equivalent dielectric-layer formulation together with the Brooks magnetic pre-sheath model, providing an engineering description of RF sheath formation within the cold-plasma approximation. Comparisons with experiments indicate that cold-plasma RF models can reproduce the spatial localization of RF-sheath activity but may overestimate the absolute rectified sheath voltages under low-density conditions where the slow wave becomes propagative. The principal conclusions of the present work therefore rely primarily on the qualitative trends in RF sheath acceleration and its coupling to local plasma conditions rather than on the exact magnitude of the peak sheath voltage.

The high- and low-density reference cases summarized in \ref{app:rfsheath} are included to illustrate the qualitative sensitivity of RF sheath formation to edge-plasma density and slow-wave accessibility. They should not be interpreted quantitatively in relation to the present RF-sheath solution because the reference calculations differ in plasma profiles, antenna excitation, absorbed power, plasma--vacuum treatment, sheath-active surfaces, and sheath formulation. The wide-grid SOLPS-ITER background employed in the present work instead provides the plasma conditions used consistently throughout the coupled STRIPE calculation. It should also be noted that the present RF sheath
model does not include RF sheath formation on remote
magnetically or electrically connected plasma-facing
components. Such long-range RF sheath interactions
have previously been discussed for JET
ICRH antennas~\cite{Klepper:2013,Klepper:2016}. While
the Petra-M calculations indicate RF-active regions
over the full antenna panel, quantifying their
contribution to RF-driven erosion in ITER would
require dedicated RF sheath calculations on those
surfaces and is beyond the scope of the present work.

Additional modeling assumptions should also be considered when interpreting the erosion and transport results. The SOLPS-ITER background treats the main plasma as pure D owing to the current modeling limitation, whereas the corresponding $Q=10$ burning-plasma scenario would nominally contain an approximately 50/50 D--T fuel mixture. This difference is not accounted for in the present erosion sampling and may affect the relative contribution of the main fuel ions to W sputtering. The impurity calculations are also performed in the trace-impurity approximation, such that the evolving W population does not modify the background plasma. The plasma density, temperature, electrostatic potential, and flow fields are prescribed from the SOLPS-ITER solution and do not include RF-induced changes in plasma convection, turbulence, or wall-bias-driven transport. Surface evolution, including roughening, morphology changes, mixed-material formation, and the long-term evolution of redeposited layers, is also neglected.

Beyond the local erosion process, the subsequent transport of sputtered W determines whether RF-driven PMI remains confined to the antenna vicinity or contributes to the wider impurity inventory. Immediately after sputtering, W atoms propagate as energetic neutrals and can travel substantial distances before ionization, contributing to the comparatively small prompt redeposition fraction near the antenna. Following ionization, the resulting W ions become strongly magnetized and are transported predominantly along magnetic field lines, producing an antenna-connected impurity reservoir extending through the SOL toward the divertor.

At $t=100$~ms, approximately 22.0\% of the mobile W inventory remaining in the GITRm plasma-density tally is located within the confined-plasma portion covered by the wide-grid SOLPS-ITER mesh. This fraction does not represent 22.0\% of the complete initially launched particle population because W particles that have already intersected material surfaces and been recorded as deposited are excluded from both the numerator and denominator. It instead describes the spatial partitioning of the mobile W inventory remaining in the plasma at the end of the 100-ms calculation.

The corresponding W concentration, $c_{\mathrm{W,ann}}=1.70\times10^{-6}$, is evaluated only over the SOLPS-covered annular confined-plasma region, $0.88\leq\psi_N\leq1.00$. It is therefore not a volume-averaged concentration over the complete ITER confined plasma. The magnetic-axis region is not included because the wide-grid SOLPS-ITER background does not provide a finite electron density there. Extrapolation to a whole-core concentration would require assumptions regarding both the electron inventory and the antenna-generated W density in the omitted inner-core region.

If the unsimulated inner-core region contains negligible antenna-generated W, inclusion of its electron inventory would reduce the corresponding whole-core concentration below the annular value reported here. Conversely, inward convection or a neoclassical impurity pinch could transport W into the omitted region and increase its local inventory. Because neither process is resolved in the present calculation, the annular concentration should be interpreted as a finite-domain transport metric rather than as a prediction of the complete ITER core W concentration.

The 100-ms calculation also does not represent a steady-state impurity inventory. The W population is initialized once from the RF-driven sputtering source and subsequently followed for a finite duration. The reported inventory therefore characterizes the transport and spatial redistribution occurring over 100~ms. If the effective confinement time of antenna-generated W exceeds the simulation duration, continuous RF operation could produce a larger accumulated W inventory. Conversely, additional loss processes acting on shorter timescales could reduce the retained inventory.

A quantitative steady-state assessment would require continuous source injection over longer simulation times together with realistic core transport and loss mechanisms. In particular, neoclassical convection, inward or outward pinch terms, turbulent impurity transport, and self-consistent coupling to the confined-plasma background may affect the long-time W distribution. The present results therefore establish that antenna-generated W can reach the confined-plasma edge within 100~ms, but they do not determine its long-time whole-core accumulation.

The RF-induced W source identified here should also be interpreted within the broader ITER W source budget. Recent ITER-wide calculations for the same wide-grid SOLPS-ITER background predict that thermal plasma--material interactions produce a gross W erosion source of approximately $6.8\times10^{21}$~atoms~s$^{-1}$ in the divertor and approximately $9.9\times10^{20}$~atoms~s$^{-1}$ in the main chamber~\cite{Baumann2026}. The RF-specific antenna source of $3.34\times10^{18}$~atoms~s$^{-1}$ is therefore approximately three orders of magnitude smaller than the divertor source and at least two orders of magnitude smaller than the main-chamber source. The ICRH antenna is consequently unlikely to dominate the overall ITER gross W source under the conditions considered here. This conclusion, however, reflects a plasma background with relatively strong Ne-driven thermal erosion. In scenarios optimized to suppress thermal W erosion, particularly through reduced Ne seeding, the relative importance of RF-driven antenna erosion may increase. Assessing the antenna contribution under such backgrounds is therefore an important direction for future work.

More broadly, the present work demonstrates the importance of treating RF wave propagation, sheath formation, charge-state-dependent ion acceleration, sputtering, and impurity transport within a unified framework. Individually, each model captures only one part of the RF-driven PMI process. Their coupled implementation within STRIPE resolves the sequence from RF wave propagation and sheath rectification to geometry-specific erosion and subsequent W migration.

These results indicate that mitigating RF-driven erosion requires more than reducing the peak RF sheath voltage. The strongest erosion occurs at surface locations where the local RF-modified IEADs are sufficiently energetic and the incident plasma flux is large. Effective mitigation strategies should therefore seek to reduce and optimize the spatial distribution of parallel RF electric fields, for example through antenna-feeding optimization, while also considering limiter geometry, magnetic-field incidence, local edge-plasma conditions, and impurity seeding. The STRIPE framework provides a predictive capability for evaluating these coupled effects and supporting the optimization of ICRH antenna designs and impurity-control strategies for ITER and future reactor-class fusion devices.

\section{Summary and Conclusions}
\label{sec:8}

This work presents the first integrated application of
the STRIPE framework to investigate the scenario-specific  RF sheath-driven
tungsten erosion and global impurity transport at the
ITER ICRH antenna under Ne-seeded operating
conditions. By coupling wide-grid
SOLPS-ITER plasma backgrounds, full-wave RF sheath
calculations in COMSOL, energy- and
angle-dependent sputtering yields from RustBCA,
geometry-specific IEADs from GITR, and
three-dimensional impurity transport using GITRm,
the present study provides the first self-consistent
prediction of RF-driven plasma--material interaction
at the ITER ICRH antenna within an ITER-scale,
all-tungsten environment.

The principal conclusions are summarized as follows:

\begin{itemize}

\item Comparison between thermal and RF sheath
conditions demonstrates that RF sheath acceleration
fundamentally modifies the tungsten source at the
ITER ICRH antenna. Under the present ITER baseline
conditions and using the upper-bound considerations, RF sheath acceleration increases the
area-integrated gross W erosion by approximately a
factor of 64 relative to the thermal sheath case.

\item The spatial distribution of RF-driven tungsten
erosion is governed by the combined effects of
RF-modified ion energy--angle distributions (IEADs)
and the local incident plasma flux rather than by the
RF sheath voltage alone. Consequently, the locations
of maximum RF sheath potential do not necessarily
coincide with the locations of maximum erosion,
demonstrating that predictive assessment of RF-driven
PMI requires coupled treatment of sheath
acceleration, particle transport, and surface
interaction physics.

\item Ne impurities remain the dominant contributors
to the primary plasma-ion sputtering source.
Although RF sheath acceleration increases the D
contribution to approximately 12.6\%, Ne still
accounts for approximately 87.4\% of the primary
plasma-ion source. The largest individual
contributions arise from Ne$^{2+}$, Ne$^{8+}$,
Ne$^{3+}$, and Ne$^{4+}$, while W self-sputtering
provides only a modest secondary contribution
($\sim5.5\%$ of the total gross erosion).

\item Approximately 10\% of the sputtered W is
locally redeposited under both thermal and RF sheath
conditions. Thus, RF sheath acceleration primarily
increases the magnitude of the antenna W source
rather than the local redeposition fraction.

\item Although RF sheath acceleration increases the
gross antenna W erosion by approximately a factor of
64 relative to the thermal sheath case, the
resulting RF-specific gross W source remains
approximately three orders of magnitude smaller than
the dominant thermal divertor source and at least two order of magnitude smaller than than from the main chamber predicted for
the same wide-grid SOLPS-ITER plasma background.
Consequently, the ITER ICRH antenna is unlikely to
dominate the overall tungsten source budget under the
operating conditions considered.

\item Global impurity transport calculations show
that sputtered W initially propagates ballistically
as neutrals before becoming ionized and transported
predominantly along antenna-connected magnetic field
lines. After 100~ms, approximately 22\% of the
mobile W inventory remaining in the plasma resides
within the SOLPS-covered confined-plasma region,
corresponding to an annular W concentration of
$c_{\mathrm{W,ann}}=1.70\times10^{-6}$ over
$0.88\leq\psi_N\leq1.00$. The predicted W density
remains several orders of magnitude below the
background plasma density, supporting the validity of
the trace-impurity approximation adopted in the
transport calculations.

\item The reported transport results correspond to a
finite-time (100~ms) evolution following a single
initialization of the RF-generated W source rather
than to a steady-state impurity inventory.
Quantitative prediction of long-time accumulation
will require continuous source injection together
with extended transport simulations and additional
core transport physics.

\item The integrated STRIPE framework demonstrates
that predictive assessment of RF-driven plasma--
material interaction requires simultaneous treatment
of RF wave propagation, sheath formation,
charge-state-dependent ion acceleration,
geometry-specific IEADs, sputtering, redeposition,
and three-dimensional impurity transport. The
framework therefore provides a predictive capability
for evaluating RF-specific tungsten sources and
supporting ITER ICRH antenna optimization and future
fusion reactor design.

\end{itemize}

Overall, the present work demonstrates that the
ITER ICRH antenna is unlikely to dominate the global
ITER tungsten source budget under the operating
conditions considered. Nevertheless, RF sheath
acceleration generates a localized main-chamber W
source whose magnitude and spatial distribution are
controlled by the coupled interaction of RF sheath
physics, geometry-specific IEADs, and incident plasma
flux. The integrated STRIPE framework developed here
provides a predictive methodology for quantifying
this RF-specific contribution and for guiding the
optimization of ICRH antenna designs and impurity
control strategies for ITER and future fusion power
plants.
\section*{Acknowledgement}
 The authors would like to thank Dr. Christopher C.  Klepper for his careful reading of the manuscript and for his valuable comments and suggestions. This material is based upon work supported by the U.S. Department of Energy, Office of Science, Office of Advanced Scientific Computing Research and Office of Fusion Energy Science, Scientific Discovery through Advanced Computing (SciDAC) program. This research used resources of the Fusion Energy Division, FFESD and the ORNL Research Cloud Infrastructure at the Oak Ridge National Laboratory, which is supported by the Office of Science of the U.S. Department of Energy under Contract No. DE-AC05-00OR22725. 
 
 The views and opinions expressed herein do not necessarily reflect those of the ITER Organization.

\appendix
\section{Sensitivity of RF Sheath Formation to ITER Edge Plasma Conditions}
\label{app:rfsheath}
%%%%%%%%%%%%%%%%%%%%%%%%%%%%%%%%%%%%%%%%%%%%%%%%%%%%%%%%%%%%%%%%%%%%%%

The RF sheath calculations presented in Section~\ref{sec:4} employ the
wide-grid SOLPS-ITER plasma background described in
Section~\ref{sec:3}. Because RF sheath formation depends sensitively on
the local edge plasma through the cold-plasma dielectric response, it is
useful to compare the present ITER-relevant plasma background with the
high- and low-density reference cases previously investigated using
COMSOL and Petra-M. Those reference calculations employed edge plasma
profiles derived from the 2010 ITER baseline equilibrium and illustrate
how changes in the edge plasma modify slow-wave accessibility and the
resulting RF sheath structure. 

\begin{figure*}
\centering

\begin{minipage}[t]{0.5\linewidth}
\centering
\textbf{(a)}\\[2pt]
\includegraphics[width=\linewidth]{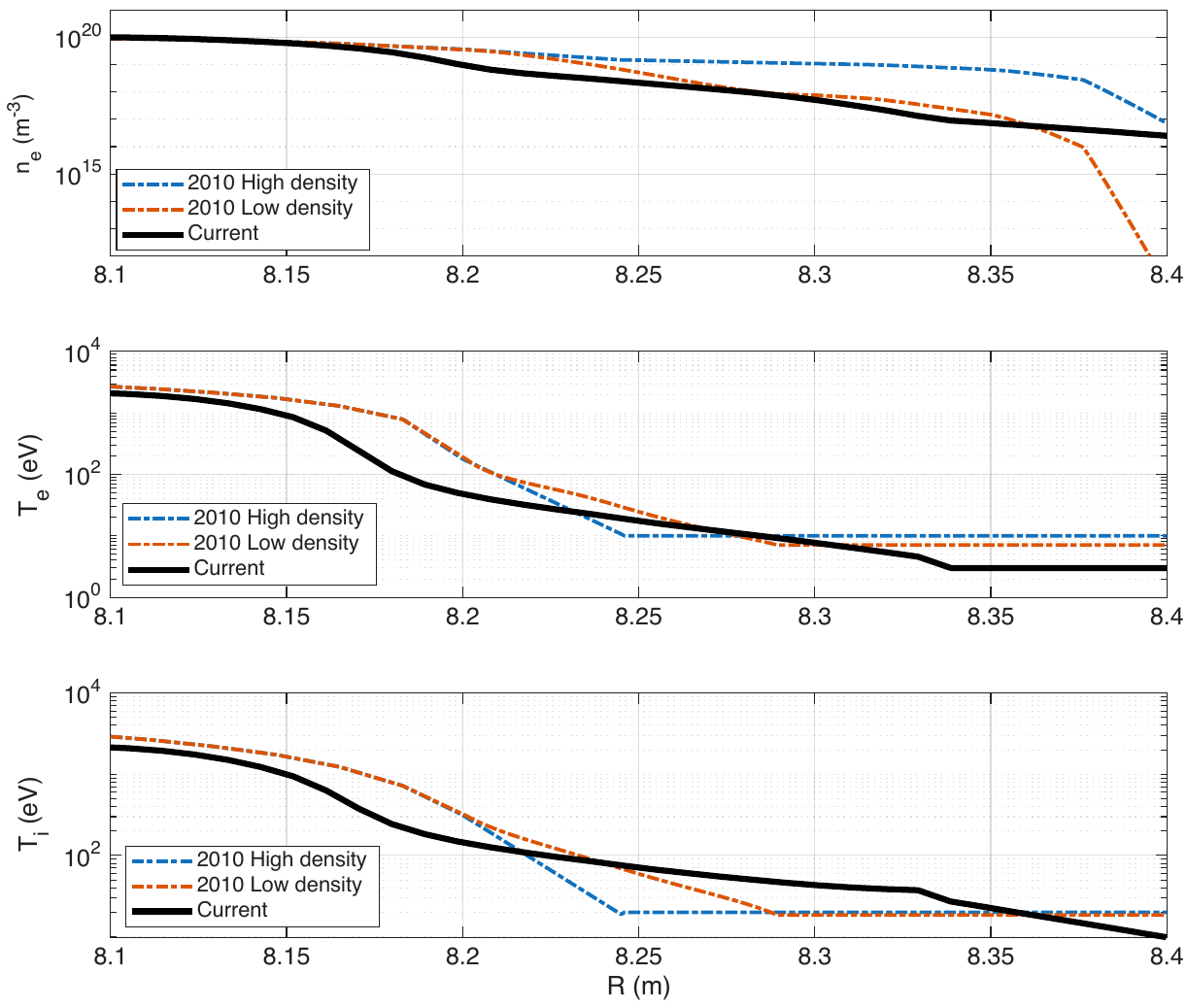}
\end{minipage}
\hfill
\begin{minipage}[t]{0.43\linewidth}
\centering
\textbf{(b)}\\[2pt]
\includegraphics[width=\linewidth]{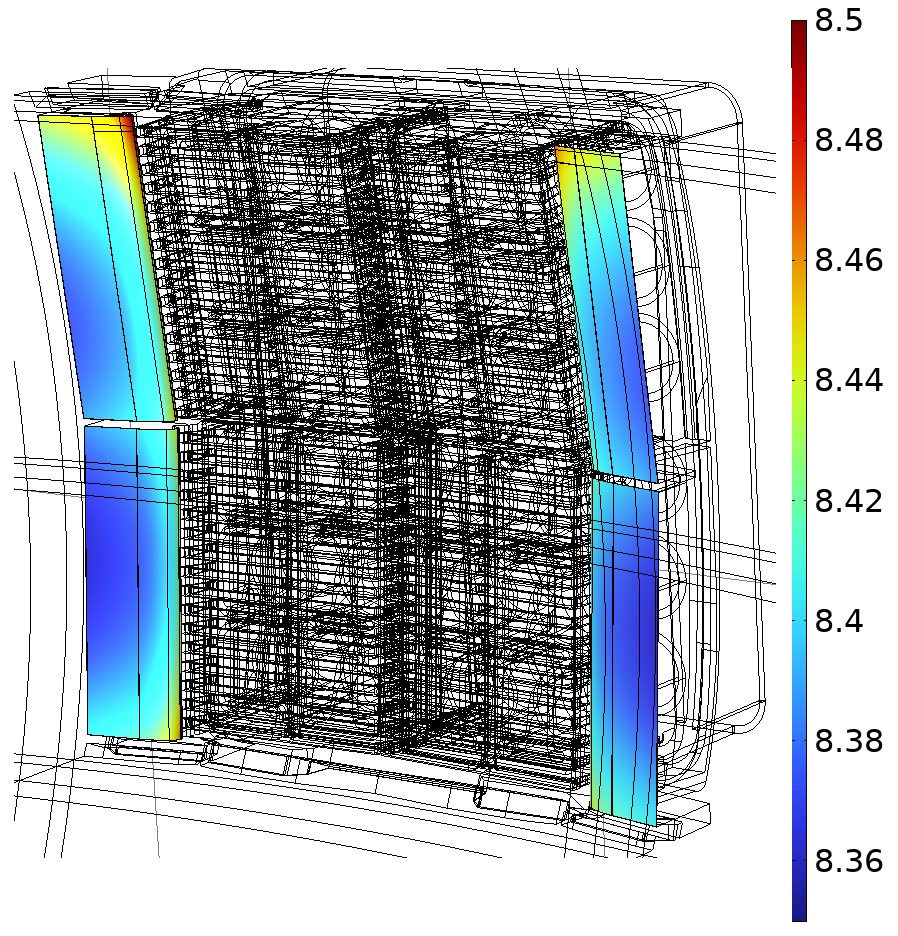}
\end{minipage}

\caption{
(a) Comparison of the reference 2010 high-density (blue), 2010
low-density (orange), and present wide-grid SOLPS-ITER (black) edge
plasma profiles, showing the electron density ($n_e$), electron
temperature ($T_e$), and ion temperature ($T_i$) along the outer
midplane. Near the ICRH antenna, the present electron density lies
between the two reference density profiles, while the temperature
profiles reflect the self-consistent SOLPS-ITER transport solution.
(b) Corresponding radial coordinate [m] employed for mapping the
one-dimensional plasma profiles onto the ITER ICRH antenna geometry.}
\label{fig:app_density}
\end{figure*}

Figure~\ref{fig:app_density}(a) compares the electron density,
electron temperature, and ion temperature profiles for the two
reference cases with the wide-grid SOLPS-ITER plasma background used in
the present work. In the vicinity of the ICRH antenna, the present
electron density lies between the 2010 high- and low-density reference
profiles, whereas the corresponding electron and ion temperature
profiles differ more substantially because they are obtained from a
self-consistent SOLPS-ITER transport solution rather than the
prescribed 2010 edge profiles. Figure~\ref{fig:app_density}(b) shows
the corresponding radial coordinate employed to map the
one-dimensional profiles onto the ITER ICRH antenna geometry.

In the low-density reference case, the plasma density decreases rapidly
inside the recessed antenna cavity, placing most limiter surfaces
within the slow-wave propagative region ($S>0$). Under these
conditions, standard cold plasma theory, and hence also the numerical results, predict resonance cone formation at resonance cone tangency
points, at which the normal electric field is severely increased even in the collisional case \cite{tierens2023resonance,tierens2024slow,tierens2026power}. Experiments confirm at least the presence of these wave modes\cite{paulus2025icrf}, though recent efforts e.g. on WEST\cite{Diab:2026} have shown that safe operation is possible even in this regime. In contrast, the high-density reference case maintains the
plasma-facing limiter surfaces predominantly within the
slow-wave-evanescent region ($S<0$), producing a qualitatively
different sheath structure characterized mainly by field-aligned
streaks. The wide-grid SOLPS-ITER plasma background employed in the
present work contains regions of both $S>0$ and $S<0$, representing a
reactor-relevant mixed operating regime used throughout the STRIPE
workflow.

\begin{figure*}
\centering
\includegraphics[width=\linewidth]{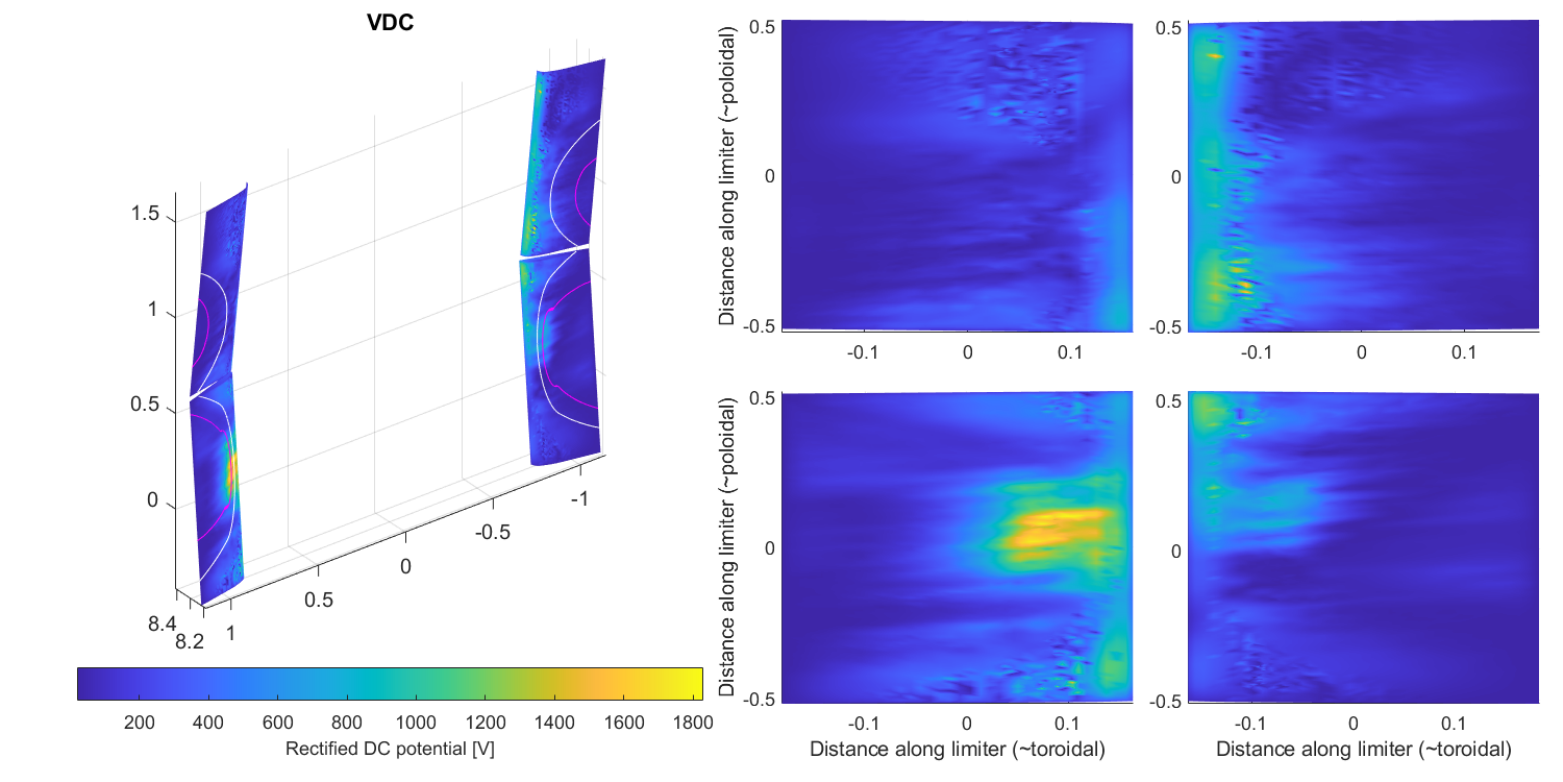}
\caption{
Rectified RF sheath potential predicted using the nonlinear COMSOL
sheath-equivalent dielectric-layer formulation for the low-density
reference case. The limiter surfaces reside predominantly within the
slow-wave propagative region ($S>0$), producing localized sheath
potentials near the resonance-cone tangency locations.
}
\label{fig:app_lowdensity}
\end{figure*}

Figure~\ref{fig:app_lowdensity} presents the RF sheath solution
predicted using the nonlinear COMSOL sheath-equivalent dielectric-layer
formulation \cite{Beers1,Tierens_2024} for the low-density reference case. Because the limiter
surfaces reside predominantly within the slow-wave propagative region
($S>0$), resonance-cone propagation produces localized kilovolt-level
RF sheath potentials near the resonance-cone tangency locations.

\begin{figure}
\centering
\includegraphics[width=0.85\linewidth]{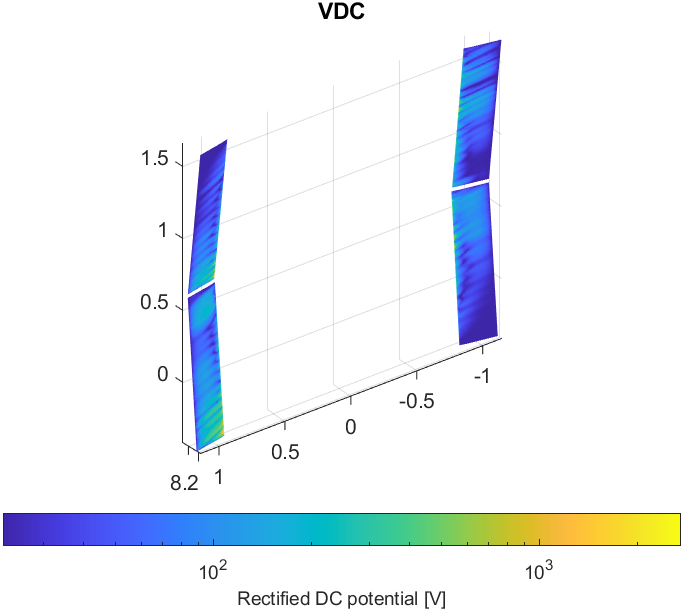}
\caption{
Rectified RF sheath potential predicted using the nonlinear COMSOL
sheath-equivalent dielectric-layer formulation for the high-density
reference case. In contrast to the low-density scenario, the sheath
structure is characterized primarily by field-aligned streaks,
reflecting the predominance of slow-wave-evanescent ($S<0$)
conditions on the plasma-facing limiter surfaces.
}
\label{fig:app_comsol_high}
\end{figure}

\begin{figure}
\centering
\includegraphics[width=0.85\linewidth]{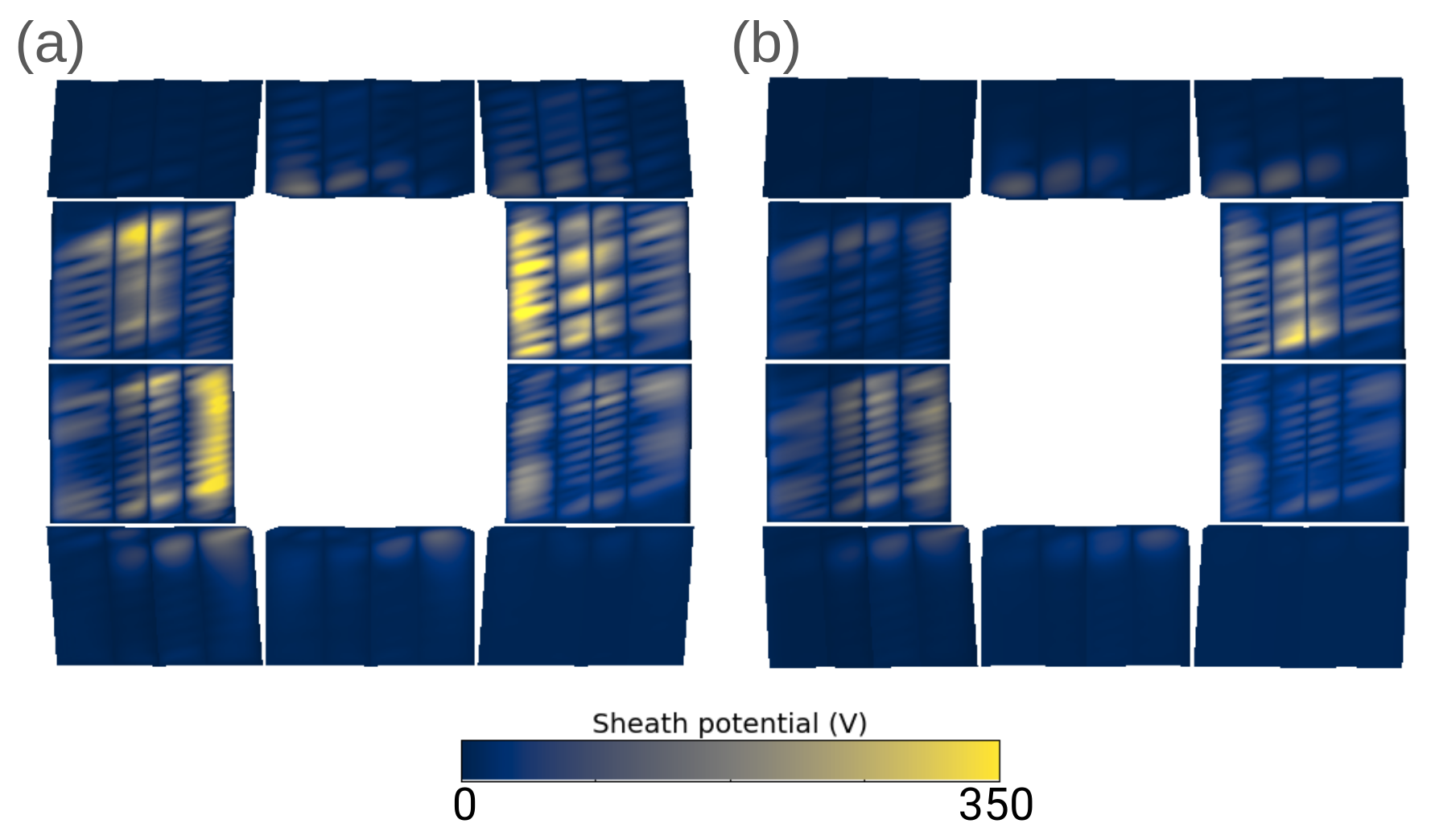}
\caption{
RF sheath potential ($V_{\rm RF}$) predicted using the
Petra-M sheath boundary-condition formulation for the
high-density reference case with (a)
$n_{\rm min}=3\times10^{17}~\mathrm{m}^{-3}$ and (b)
$n_{\rm min}=1\times10^{18}~\mathrm{m}^{-3}$. Both cases
exhibit field-aligned RF-active streaks on the ITER ICRH
antenna limiter surfaces, characteristic of the
slow-wave-evanescent ($S<0$) regime. Although the
absolute sheath potentials differ from the COMSOL
results because of the different modeling approaches,
both calculations identify similar RF-active regions on
the limiter surfaces.
}
\label{fig:app_petram_high}
\end{figure}

The corresponding high-density COMSOL result is shown in
Fig.~\ref{fig:app_comsol_high}, while the Petra-M prediction for the
same reference plasma is presented in
Fig.~\ref{fig:app_petram_high}. Both calculations identify similar
RF-active regions on the ITER antenna limiters and demonstrate the
transition from resonance-cone-dominated localization in the
low-density ($S>0$) regime to predominantly field-aligned sheath
structures in the high-density ($S<0$) regime.

The absolute sheath voltages predicted by COMSOL and Petra-M differ for the cases considered here. Although previous benchmark studies  have shown good qualitative agreement between the two approaches when applied to comparable  ITER plasma parameters and ICRH antenna conditions~\cite{Bertelli:2023}, the present calculations are not intended as a direct code-to-code comparison and do not employ identical simulation conditions. The models differ in the antenna excitation, absorbed power, plasma--vacuum treatment, sheath-active surfaces, and sheath formulation. COMSOL employs an iteratively coupled nonlinear sheath-equivalent dielectric-layer model~\cite{Tierens_2024}, whereas Petra-M uses an asymptotic sheath boundary-condition formulation~\cite{Shiraiwa:2017}. Consequently, the peak sheath voltages summarized in Table~\ref{tab:app_validation} should not be interpreted as a direct quantitative comparison between the two approaches. Rather, the comparison demonstrates the consistent qualitative localization of RF-active regions and illustrates the sensitivity of RF sheath formation to edge plasma density and slow-wave accessibility. The wide-grid SOLPS--ITER plasma background adopted in the present work provides the ITER-relevant plasma conditions used throughout the STRIPE workflow for the subsequent tungsten erosion and impurity transport calculations.

\begin{table*}
\caption{
Representative ITER ICRH RF sheath calculations. The COMSOL and
Petra-M calculations employ different plasma scenarios, absorbed
powers, plasma--vacuum treatments, sheath-active surfaces, and sheath
formulations; therefore, the listed peak sheath voltages should not be
interpreted as a direct quantitative code comparison.
}
\centering
\begin{tabular}{lcccc}
\hline
Case & Plasma regime & Power & Peak $V_{\rm RF,DC}$ & Dominant sheath characteristic \\
\hline
COMSOL & Low density &
$\sim9.5$ MW absorbed &
$\sim1.8$ kV &
Resonance-cone localization ($S>0$) \\

COMSOL & High density &
$\sim0.55$ MW absorbed &
$\sim2.7$ kV &
Field-aligned limiter streaks ($S<0$) \\

Petra-M & High density &
$\sim1.3$ MW absorbed &
$\sim210$ V &
Field-aligned RF-active regions ($S<0$) \\

Present work &
Wide-grid SOLPS-ITER &
7 MW coupled &
$\sim3$ kV &
Mixed $S>0$/$S<0$ accessibility \\
\hline
\end{tabular}
\label{tab:app_validation}
\end{table*}

\section{Benchmark of the GITR Thermal Sheath Model}
\label{app:gitr_validation}

The erosion calculations presented in this work use
the GITR implementation of the Brooks magnetic
pre-sheath formulation to describe ion acceleration
through the combined Debye sheath and magnetic
pre-sheath before impact on plasma-facing components.
Because the resulting IEADs directly determine the geometry-specific
effective sputtering yields, the underlying sheath
electric-field and particle-trajectory models are
benchmarked independently of the RF sheath
calculations.

The GITR thermal sheath model is benchmarked with the
analytical formulation developed by Borodkina
\textit{et al.}~\cite{Borodkina:2015,Borodkina:2016}.
The Borodkina model provides analytical descriptions
of the near-wall potential distribution and ion
trajectories for an obliquely incident magnetic field
and has previously shown good agreement with fluid
solutions and particle-in-cell simulations. The
present comparison considers thermal sheath conditions
only and therefore validates the near-wall
electric-field and particle-acceleration model
implemented in GITR, rather than the rectified RF
sheath potentials computed using COMSOL.

\begin{figure}
\centering
\includegraphics[width=\linewidth]
{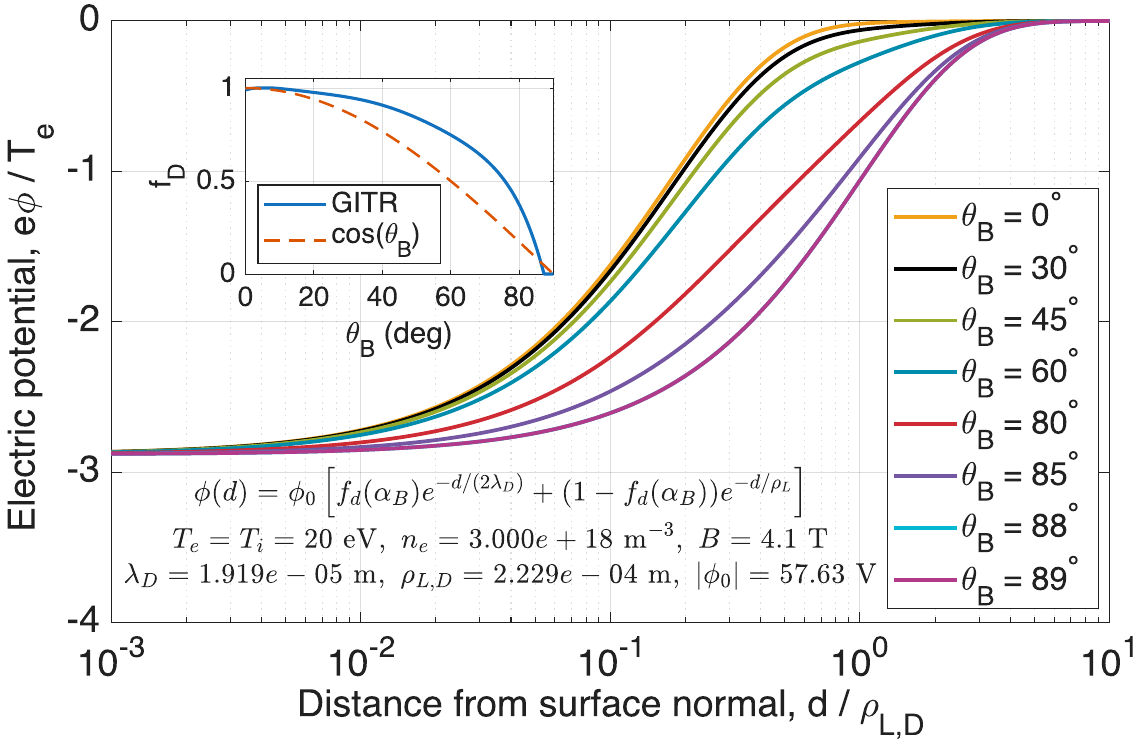}
\caption{
GITR/GITRm thermal sheath model obtained using the
electric-field formulation defined in Eqs.~(4)
and~(5). The normalized electrostatic potential,
$e\phi/T_e$, is shown as a function of the
wall-normal distance normalized to the deuterium
Larmor radius, $d/\rho_{L,D}$, for magnetic-field
incidence angles between $\theta_B=0^\circ$ and
$89^\circ$. The inset shows the Debye-sheath
potential fraction, $f_D(\theta_B)$, used to
partition the total sheath potential between the
Debye sheath and magnetic pre-sheath. The
representative plasma conditions are
$T_e=T_i=20~\mathrm{eV}$,
$n_e=3\times10^{18}~\mathrm{m}^{-3}$, and
$B=4.1~\mathrm{T}$.
}
\label{fig:gitr_sheath_model}
\end{figure}

Figure~\ref{fig:gitr_sheath_model} illustrates the
thermal sheath formulation implemented in GITR. As
defined in Eq.~\ref{eq:brooks_partition} and Eq.~\ref{eq:brooks_field}, the total sheath
potential is partitioned between the Debye sheath and
the magnetic pre-sheath according to the local
magnetic-field incidence angle. The corresponding
normal electric field is represented by the sum of a
rapidly decaying Debye-sheath contribution with
characteristic scale $2\lambda_D$ and a more extended
magnetic-pre-sheath contribution with characteristic
scale $\rho_i$. No additional sheath-potential
expression is introduced here; the profiles shown in
Fig.~\ref{fig:gitr_sheath_model} are obtained directly
from the formulation defined in the main text.

For nearly normal magnetic-field incidence, most of
the potential drop occurs within the Debye sheath,
and the potential approaches the plasma value over a
relatively short distance. As the magnetic field
becomes increasingly tangential to the surface,
$f_D$ decreases and a progressively larger fraction
of the potential drop is transferred to the magnetic
pre-sheath. The potential profile therefore extends
over distances comparable to the ion Larmor radius,
modifying both the ion acceleration history and the
resulting surface-impact distribution.

\begin{figure*}
\centering
\includegraphics[width=\textwidth]
{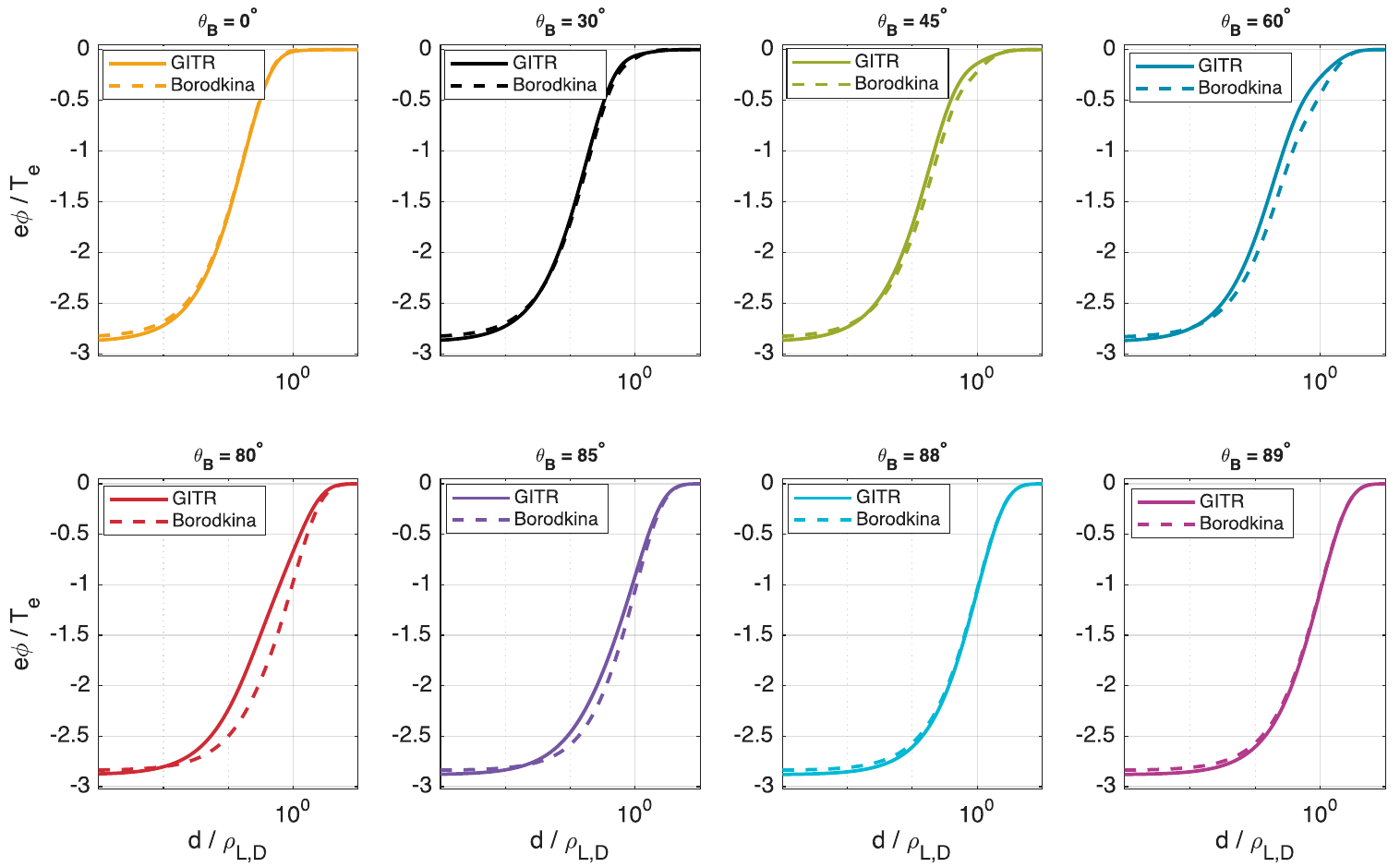}
\caption{
Comparison of normalized electrostatic potential
profiles predicted by the GITR thermal sheath model
(solid lines) and the analytical model of Borodkina
\textit{et al.} (dashed lines) for magnetic-field
incidence angles between $\theta_B=0^\circ$ and
$89^\circ$. The GITR profiles are calculated using
the sheath electric-field formulation defined in
Eqs.~(4) and~(5). The wall-normal distance is
normalized to the deuterium Larmor radius. This
comparison applies only to thermal sheath conditions.
}
\label{fig:gitr_potential_validation}
\end{figure*}

Figure~\ref{fig:gitr_potential_validation} compares
the normalized potential profiles obtained from the
GITR implementation with those predicted by the
Borodkina analytical model for magnetic-field
incidence angles ranging from normal incidence,
$\theta_B=0^\circ$, to nearly grazing incidence,
$\theta_B=89^\circ$.

The two models exhibit close agreement over the full
range of magnetic-field incidence angles considered.
GITR reproduces both the characteristic potential
variation near the surface and the progressive
extension of the magnetic pre-sheath as the magnetic
field becomes increasingly grazing. The agreement is
particularly close for normal and moderately inclined
magnetic fields. Relatively small differences develop
at intermediate grazing angles, but the overall
potential magnitude, spatial extent, and
magnetic-field-angle dependence remain consistent
with the analytical formulation.

A more direct test of the IEAD calculation is obtained
by comparing the resulting ion impact-angle
distributions. These distributions depend on the
combined effects of the sheath electric field,
magnetic-field geometry, initial ion velocity
distribution, gyromotion, and particle integration
through the near-wall region.

\begin{figure*}
\centering
\includegraphics[width=\textwidth]
{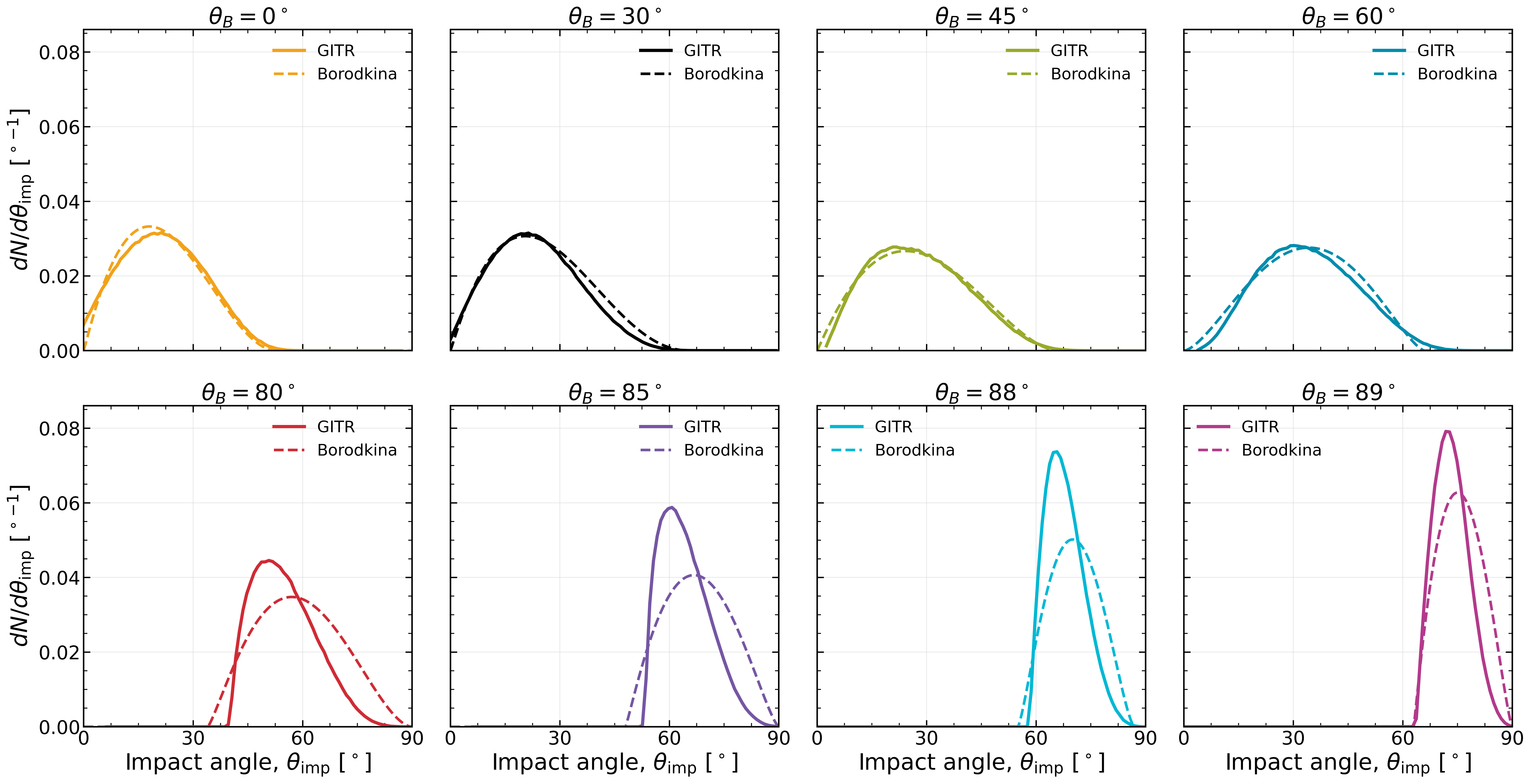}
\caption{
Comparison of ion impact-angle distributions predicted
by the GITR thermal sheath model (solid lines) and the
analytical particle-trajectory model of Borodkina
\textit{et al.} (dashed lines) for magnetic-field
incidence angles between $\theta_B=0^\circ$ and
$89^\circ$. The impact angle,
$\theta_{\rm imp}$, is measured relative to the
surface normal, and each distribution is normalized
such that its integral over impact angle is unity.
This comparison applies only to thermal sheath
conditions.
}
\label{fig:gitr_angle_validation}
\end{figure*}

As shown in
Fig.~\ref{fig:gitr_angle_validation}, the GITR and
Borodkina impact-angle distributions agree closely
for normal and moderately inclined magnetic fields
($\theta_B\leq60^\circ$). Both models predict that
the most probable impact angle shifts toward more
grazing incidence as the magnetic field becomes
increasingly tangential to the surface. For
$\theta_B\geq80^\circ$, GITR reproduces the same
overall shift and narrowing of the impact-angle
distribution, with only small differences in the peak
magnitude and distribution width. These differences
become more noticeable at the largest magnetic-field
incidence angles, where near-wall ion dynamics are
particularly sensitive to the partitioning of the
sheath potential between the Debye sheath and the
magnetic pre-sheath, as well as to the assumed ion
distribution at the sheath entrance.

Taken together,
Figs.~\ref{fig:gitr_potential_validation} and
\ref{fig:gitr_angle_validation} demonstrate that the
GITR implementation reproduces the principal
potential-profile and ion-impact-angle dependencies of
the Borodkina analytical thermal sheath model. The
benchmark therefore provides confidence in the
near-wall particle acceleration and magnetic
pre-sheath treatment used to generate the
geometry-specific IEADs employed in the present
erosion calculations.

This benchmark is restricted to thermal sheath
conditions. For the RF calculations presented in the
main text, the total rectified sheath potential is
obtained from the nonlinear COMSOL RF sheath
solution. GITR then applies the same benchmarked
near-wall particle-acceleration framework defined in
Eq.~\ref{eq:brooks_partition} and Eq.~\ref{eq:brooks_field} to distribute the prescribed
potential between the Debye sheath and magnetic
pre-sheath according to the local magnetic-field
incidence angle and to compute the corresponding
charge-state-dependent IEADs.
%%%%%%%%%%%%%%%%%%%%%%%%%%%%%%%%%%%%%%%%%%%%%%%%%%%%%%%%%%%%%%%%%%%%%%%%%%

\section{Global Tungsten Transport with GITRm}
\label{app:gitrm}

GITRm~\cite{nath:2023,Nath:2025} is a three-dimensional unstructured-mesh Monte Carlo particle-tracking code developed for impurity transport, plasma--material interaction, material erosion, and re-deposition in magnetically confined fusion plasmas. Built on the PUMIPic framework~\cite{Diamond:2021}, GITRm employs fully three-dimensional tetrahedral meshes, allowing realistic representation of complex plasma-facing component (PFC) geometries such as the ITER ICRH antenna. The present calculations use the trace-impurity approximation, in which sputtered W impurities evolve within prescribed plasma and electromagnetic fields without modifying the background plasma. This approximation is appropriate because the tungsten density remains much smaller than the background plasma density throughout the simulations.

\subsection{Transport Model}

The impurity distribution function,
$f_z(\mathbf{r},\mathbf{v},t)$,
is evolved using the steady-state Boltzmann transport equation,

\begin{equation}
\mathbf{v}_z\cdot\nabla f_z
+
\frac{q_z}{m_z}
\left(
\mathbf{E}
+
\mathbf{v}_z\times\mathbf{B}
\right)
\cdot
\frac{\partial f_z}
{\partial\mathbf{v}_z}
=
C_{zb}
+
S_{\rm sput}
+
S_{\rm self}
+
S_{\rm at},
\label{eq:gitrm_boltzmann}
\end{equation}

where $q_z$ and $m_z$ denote the impurity charge and mass,
$C_{zb}$ represents Coulomb collisions with the background plasma, and
$S_{\rm sput}$,
$S_{\rm self}$,
and
$S_{\rm at}$
represent primary sputtering, W self-sputtering, and atomic processes, respectively.

Particle trajectories are advanced by integrating the Lorentz force,

\begin{equation}
\mathbf{F}_{\rm L}
=
q_z
\left(
\mathbf{E}
+
\mathbf{v}_z\times\mathbf{B}
\right),
\label{eq:lorentz}
\end{equation}

together with collisional slowing down,

\begin{equation}
\mathbf{F}_{\rm drag}
=
-\nu_s m_z
\left(
\mathbf{v}_z
-
\mathbf{v}_{\rm flow}
\right),
\label{eq:drag}
\end{equation}

where $\mathbf{v}_{\rm flow}$ is obtained from the SOLPS-ITER plasma background. The electric field includes both the background plasma field and the RF sheath field imported from COMSOL, naturally incorporating RF sheath acceleration together with the associated
$\mathbf{E}\times\mathbf{B}$
drift.

The general GITRm formulation also permits inclusion of parallel thermal forces,

\begin{equation}
\mathbf{F}_{\nabla T}
=
\alpha_e
\nabla_\parallel(k_BT_e)
+
\beta_i
\nabla_\parallel(k_BT_i),
\label{eq:thermal}
\end{equation}

where

\begin{equation}
\nabla_\parallel
=
\mathbf{b}
(\mathbf{b}\cdot\nabla),
\qquad
\mathbf{b}
=
\frac{\mathbf{B}}
{|\mathbf{B}|}.
\end{equation}

For the present ITER simulations, which are restricted to the upstream ICRH antenna region, the parallel ion-temperature gradients are comparatively weak. Consequently, the thermal-force contribution remains small relative to Lorentz acceleration and collisional slowing down and is therefore neglected. This approximation is not expected to remain valid in regions with strong parallel temperature gradients, such as the divertor

Cross-field turbulent transport is represented through anomalous diffusion,

\begin{equation}
\mathbf{r}
=
\mathbf{r}_0
+
\sqrt{4D_\perp\Delta t}
\,
\hat{\mathbf{b}}_\perp,
\label{eq:anom}
\end{equation}

where $D_\perp$ is the prescribed anomalous diffusion coefficient.

Ionization and recombination are treated using Maxwellian-averaged ADAS reaction-rate coefficients. For a reaction rate coefficient
$\langle\sigma v\rangle$,

\begin{equation}
\tau_{\rm at}
=
\frac{1}
{\langle\sigma v\rangle n},
\end{equation}

with reaction probability

\begin{equation}
P_{\rm at}
=
1-
\exp
\left(
-\frac{\Delta t}
{\tau_{\rm at}}
\right).
\end{equation}

These processes determine the charge-state evolution and transport of sputtered tungsten throughout the ITER edge plasma.

\subsection{Mesh, Plasma Mapping, and RF Sheath Coupling}

The computational domain consists of a fully three-dimensional unstructured tetrahedral mesh representing the ITER first wall, ICRH antenna, limiter structures, and surrounding PFCs. Local refinement is applied around the antenna aperture, FS bars, and limiter sidewalls, while a coarser mesh is employed elsewhere to reduce computational cost.

The plasma background is obtained from the wide-grid SOLPS-ITER solution described in Section~\ref{sec:3}. Following the extrapolation procedure of Section~\ref{sec:3.2}, continuous distributions of
$n_e$,
$T_e$,
$T_i$,
plasma flow,
electric potential,
and Ne charge-state densities are mapped onto the GITRm mesh assuming toroidal symmetry. Plasma quantities are interpolated to particle positions using linear finite-element interpolation.

The COMSOL-computed RF sheath potentials are mapped directly onto the antenna-facing surfaces. During particle integration, the local sheath electric field is evaluated using the Brooks sheath model together with the imported RF sheath potential, allowing particle trajectories to respond self-consistently to charge-dependent RF acceleration and magnetic-field incidence angle.

\subsection{Particle Initialization and Numerical Configuration}

Tungsten particles are initialized on the ITER ICRH limiter surfaces according to the spatially resolved gross erosion source obtained from STRIPE. The number of particles emitted from each surface facet is proportional to the local gross erosion flux and facet area, while initial energies and emission angles are sampled from a Thompson energy distribution and cosine angular distribution.

Particles subsequently undergo Lorentz acceleration, Coulomb collisions, ionization, recombination, anomalous diffusion, and repeated surface interactions. Upon striking a PFC, the impact location, charge state, energy, impact angle, and statistical weight are recorded to evaluate local redeposition and W self-sputtering. Net erosion is calculated from

\begin{equation}
\Gamma_{\rm net}
=
\Gamma_{\rm gross}
-
\Gamma_{\rm dep}.
\end{equation}

Approximately $5\times10^{6}$ computational particles
are followed using a timestep of
$\Delta t=1\times10^{-9}$~s for approximately
$100\times10^{6}$ timesteps ($\sim100$~ms physical
time). The 100~ms transport interval was selected to
characterize the finite-time migration of
antenna-generated W while maintaining computational
tractability for the full three-dimensional ITER
geometry. Throughout the simulations, a constant
anomalous tungsten diffusion coefficient of
$D_\perp=1~\mathrm{m^2\,s^{-1}}$ is employed.
Statistical convergence is verified using the time
histories of the area-integrated gross erosion, gross
deposition, net erosion, and total tungsten
inventory. Although the local gross erosion, gross
deposition, and net erosion rates stabilize rapidly
following the initial transient, the global tungsten
inventory continues to evolve because of ongoing
transport and redistribution throughout the ITER
edge plasma. Consequently, the reported 100~ms
results represent a finite-time transport assessment
rather than a converged steady-state impurity
inventory.

Simulations were performed using the
distributed-memory implementation of GITRm on four
NVIDIA A100 GPUs (80 GB per GPU). Spatial domain
decomposition and MPI particle communication enable
efficient simulation of millions of impurity
trajectories within the full three-dimensional ITER
geometry. A single 100~ms transport calculation
requires approximately one month of wall-clock time
on this computing configuration.
Under the present ITER conditions, sputtered W atoms initially propagate ballistically as neutrals until ionization. After ionization, the resulting W ions become strongly magnetized and are transported predominantly along magnetic field lines, producing local redeposition while the remaining net erosion provides the source term for the global transport calculations presented in Section~\ref{sec:6}.

\section*{References}
\bibliographystyle{ieeetr}  
\bibliography{iter}
\end{document}